\pdfoutput=1
\documentclass[fleqn,10pt]{wlscirep2}
\usepackage[utf8]{inputenc}
\usepackage[T1]{fontenc}
\usepackage{bm}
\usepackage{braket}
\usepackage{booktabs}
\usepackage{multirow}
\usepackage{graphicx}
\usepackage{listings}
\usepackage{xcolor}
\usepackage{wrapfig}
\usepackage{parskip}

\definecolor{codegreen}{rgb}{0,0.6,0}
\definecolor{codegray}{rgb}{0.5,0.5,0.5}
\definecolor{codepurple}{rgb}{0.58,0,0.82}
\definecolor{backcolour}{rgb}{0.95,0.95,0.92}

\lstdefinestyle{mystyle}{
    backgroundcolor=\color{backcolour},   
    commentstyle=\color{codegreen},
    keywordstyle=\color{magenta},
    numberstyle=\tiny\color{codegray},
    stringstyle=\color{codepurple},
    basicstyle=\ttfamily\footnotesize,
    breakatwhitespace=false,         
    breaklines=true,                 
    captionpos=b,                    
    keepspaces=true,                 
    numbers=left,                    
    numbersep=5pt,                  
    showspaces=false,                
    showstringspaces=false,
    showtabs=false,                  
    tabsize=2
}

\usepackage[final]{pdfpages}

\title{Supporting Information \\ Photovoltaics-driven power production can support human exploration on Mars}

\author[*,1,2]{Anthony J. Abel} 
\author[*,1,3]{Aaron J. Berliner} 
\author[1,4]{Mia Mirkovic}
\author[**,5,6]{William D. Collins}
\author[**,1,3,7]{Adam P. Arkin}
\author[**,1,2,8]{Douglas S. Clark}

\affil[*]{Authors contributed equally to this work}
\affil[**]{Correspondence should be addressed to W.D.C (\url{wdcollins@lbl.gov}), A.P.A. (\url{aparkin@lbl.gov}) or to D.S.C. (\url{dsc@berkeley.edu})}
\affil[1]{Center for the Utilization of Biological Engineering in Space (CUBES)}
\affil[2]{Department of Chemical and Biomolecular Engineering, University of California, Berkeley, CA 94720, USA}
\affil[3]{Department of Bioengineering, University of California, Berkeley, CA 94720, USA}
\affil[4]{Department of Electrical Engineering and Computer Sciences, University of California, Berkeley, CA 94720, USA}
\affil[5]{Lawrence Berkeley National Laboratory, Climate and Ecosystems Sciences Division, One Cyclotron Road, Berkeley, CA 94720, USA}
\affil[6]{Department of Earth and Planetary Sciences, University of California, Berkeley, CA 94720, USA}
\affil[7]{Environmental Genomics and Systems Biology Division, Lawrence Berkeley National Laboratory, One Cyclotron Road, Berkeley, 94720, CA, USA}
\affil[8]{Molecular Biophysics and Integrated Bioimaging Division, Lawrence Berkeley National Laboratory, One Cyclotron Road, Berkeley, CA 94720, USA}

\begin{abstract}
 
\end{abstract}
\begin{document}



**Title**
Photovoltaics-driven power production can support human exploration on Mars

**Authors**: Anthony J. Abel,[†,1,2] Aaron J. Berliner,[†,1,3] Mia Mirkovic,[1,4] William D. Collins,[*,5,6] Adam P. Arkin,[*,1,3,7] and Douglas S. Clark[*,1,2,8]

**Affiliations:**
1. Center for the Utilization of Biological Engineering in Space (CUBES), Berkeley, CA 94720, USA
2. Department of Chemical and Biomolecular Engineering, University of California, Berkeley, CA 94720, USA
3. Department of Bioengineering, University of California, Berkeley, CA 94720, USA
4. Department of Electrical Engineering and Computer Sciences, University of California, Berkeley, CA 94720, USA
5. Climate and Ecosystems Sciences Division, Lawrence Berkeley National Laboratory, Berkeley, CA 94720, USA
6. Department of Earth and Planetary Sciences, University of California, Berkeley, CA 94720, USA
7. Environmental Genomics and Systems Biology Division, Lawrence Berkeley National Laboratory, Berkeley, CA 94720, USA
8. Molecular Biophysics and Integrated Bioimaging Division, Lawrence Berkeley National Laboratory, Berkeley, CA 94720, USA

†These authors contributed equally to this work.
*Correspondence should be addressed to DSC (dsc@berkeley.edu), APA (aparkin@lbl.gov), or WDC (wdcollins@lbl.gov).

**Abstract**
A central question surrounding possible human exploration of Mars is whether crewed missions can be supported by available technologies using *in situ* resources. Here, we show that photovoltaics-based power systems would be adequate and practical to sustain a crewed outpost for an extended period over a large fraction of the planet's surface. Climate data were integrated into a radiative transfer model to predict spectrally-resolved solar flux across the Martian surface. This informed detailed balance calculations for solar cell devices that identified optimal bandgap combinations for maximizing production capacity over a Martian year. We then quantified power systems, manufacturing, and agricultural demands for a six-person mission, which revealed that photovoltaics-based power generation would require <10 t of carry-along mass, outperforming alternatives over ~50% of Mars' surface.

**Main Text**
Long-duration space missions or continuously-occupied extraterrestrial outposts require Earth-independent power and chemical supply. Mars has an abundance of *in situ* resources, including (sub)surface water ice (*1*) and carbon and nitrogen in atmospheric $CO_2$ and $N_2$ (*2*). Efficient conversion of these resources to reduced forms of hydrogen, nitrogen, and carbon would represent an enabling step towards sustaining a permanent human presence in space. In analogy to the proposed terrestrial "Hydrogen Economy", molecular hydrogen ($H_2$) can be used as a platform molecule for energy storage, on-demand power supply, and as a reactant driving $CO_2$ and $N_2$ (bio)chemical reduction on Mars (*3–5*).

Water electrolysis with selective catalysts can drive water reduction to $H_2$ on cathode surfaces. This technology is attractive for space manufacturing applications since reactions can proceed at high rates at

room temperature, enabling the use of low-weight, 3D-printable plastic reactors (*4*). Commercial electrolyzers can evolve $H_2$ from water with up to ~80% energy efficiency (*6*). Directly solar-powered (*i.e.*, photoelectrochemical) devices have also received significant attention, with solar-to-chemical efficiencies reaching >19% for $H_2$ production (*7*). Once generated, $H_2$ can drive $N_2$ reduction to ammonia *via* the Haber-Bosch process for crop fertilizer (*5*), $CO_2$ reduction to $CH_4$ *via* the Sabatier process or methanogenesis for ascent propellant generation (*8*), and $CO_2$ reduction to bioplastics following a variety of metabolic processes for habitat and spare parts manufacturing (*4*, *9*).

The primary alternatives for powering life support systems and chemical production facilities on Mars are miniaturized nuclear fission reactors (*10*) and photovoltaic (PV) arrays. While fission reactors are expected to behave similarly regardless of their location, the productivity limits of PV and photoelectrochemical (PEC) devices are not well-characterized for the Martian surface mainly due to differences in the surface temperature and solar intensity and spectrum from typical conditions on Earth or in space.

In an effort to determine the potential of PV and PEC devices to support a crewed mission to Mars, we integrated relevant climate data from the Mars Climate Database (*11*) into a radiative transfer model, libRadtran (*12*), to predict spectrally-resolved solar flux across the Martian surface over the course of a year. The modeling overview and sample calculations for Jezero Crater are provided in Fig. 1. Sunlight incident on the surface originating from the top of the atmosphere (TOA) is mediated by orbital geometry and local atmospheric composition of gases, ice, and dust for a given location (Fig. 1A). We determined the partial pressures of constituent gases (Fig. 1B) and the concentrations and effective radii of ice (Fig. 1C) and dust (Fig. 1D) particles as a function of altitude above the surface and provided these data as inputs to a downstream radiative transfer model (diagrammed in Fig. 1E). We then calculated the spectrally-resolved solar flux (Fig. 1F). At short wavelengths (<400 nm), light transmission through the atmosphere is limited by molecular scattering (primarily by $CO_2$) and scattering from dust particles (*13*). Scattering and absorption by gas molecules is significant at wavelengths below 300 nm, but this region is not considered here because it represents a very small fraction of the available solar flux (<0.5%). Above 400 nm, most transmission loss is due to scattering from dust particles. This is markedly different from the case on Earth, where significant molecular absorption by water molecules limits the transmission of near-infrared light.

The modeling results were used to inform efficiency calculations for PV and PEC devices producing electricity and $H_2$. Detailed balance calculations (section 4 in the Supplementary Information) (*14*, *15*) revealed ideal current-voltage characteristics for optically-thick devices consisting of 1-, 2-, and 3- junction PV and 1- and 2-junction PEC absorbers dependent on the bandgaps associated with each absorber (Fig. 2). Absorber numbers were selected to represent historical choices for PV devices on Martian rovers (*16*, *17*) and state-of-the-art PEC devices (*7*, *18*, *19*). For PEC devices, we assumed an electrical load consisting of the thermodynamic redox potential and a variable overvoltage term that incorporates loss mechanisms inevitable to a practical PEC device beyond radiative recombination already considered in the detailed balance (*14*, *15*).

The maximum efficiency for PV devices increases from 31.4% (1-junction; $E_g$=1.23 eV) to 51.3% (3-junction; $E_{g,1}$ = 1.77 eV, $E_{g,2}$ = 1.16 eV, $E_{g,3}$ = 0.72 eV) with judicious choice of bandgaps (Fig. 2A-C). For PEC devices, optimal bandgap choice and efficiency are strongly dependent on system losses (Fig. 2D-F), reflecting the importance of careful device construction and catalyst selection (*20*). For a realistic overvoltage loss of 700 mV (*14*, *15*, *20*), a maximum solar-to-chemical efficiency of 27.8% is feasible for $H_2$ production.

To evaluate the potential for solar cells to supply power and commodity chemicals, we determined the maximum practical production capacity for 3-junction PV (operating at 80% of the detailed balance limit) and 2-junction PEC devices (with a 700-mV overvoltage) over the course of a Martian year (Fig. 3). Daily and seasonal variation in solar flux and temperature (Fig. 3A, B) cause substantial (~27% deviation from the yearly average) changes in production rates (Fig. 3C, D). We defined solar day (sol) 0 at a solar longitude (Ls) of 0° (vernal equinox) and assumed the solar cell operating temperature was equal to the surface temperature at all points. Dust storm season begins at sol ~372 (Ls~180°) and is primarily responsible for the drop in production capacity from a peak of ~1.7 kWh/m$^2$/day at Jezero Crater to a minimum of ~1.0 kWh/m$^2$/day at the height of dust storm intensity around the winter solstice (Ls ~270°, sol ~514).

Bandgap combinations that maximize production over the course of a year are 5-15% different from those that optimize efficiency at solar noon (Table S7). For both PV and PEC devices, the top junction bandgap shifted up (for $H_2$-generating PEC devices, from 1.64 eV to 1.77 eV), while the bottom junction bandgap shifted down (from 0.95 eV to 0.83 eV). Hence, the photon absorption window for the bottom junction is broadened (by ~35% for the PEC device). This likely works to maximize productivity during the less dusty season (higher solar flux, Fig. 3B) by accounting for the relative blue-shift of surface-incident light (Fig. 1F) due to reduced scattering.

Production capacity of power and commodity chemicals must compare favorably to the demand necessary to sustain a Martian habitat and depends on the outpost location on the planet surface (Fig. 4A). Moreover, energy storage capacity is crucial for solar-powered production systems because the sun sets daily. We therefore developed a detailed process model to account for power systems demands, including habitat maintenance (for example, habitat temperature control and pressurization), fertilizer production for agriculture, methane production for ascent propellant, and bioplastics production for spare parts manufacturing (Fig S12). We considered four different power generation scenarios: (1) nuclear power generation with the miniaturized nuclear fission Kilopower system; (2) PV power generation with battery energy storage (PV+B); (3) PV power generation with compressed $H_2$ energy storage produced via electrolysis (PV+E); and (4) PEC $H_2$ generation with compressed $H_2$ energy storage (PEC). In our calculations, we assumed a capacity factor of 75% to account for the solar flux deviation throughout the Martian year (Fig. 3) and sized energy storage systems (batteries or compressed $H_2$) to enable 1 full day of operations from reserve power. We then calculated the carry-along mass requirements for each of the power generation systems considered.

Of the three solar-driven power generation options, only the PV+E system outcompetes the nuclear system based on carry-along mass (Fig. 4B, C; supplementary Fig. S13). For the PV+E system, the total carry-along mass increases from ~8.3 t near the equator to ~22.4 t near the South Pole (Fig. 4B), corresponding to the reduced average daily power generation of the PV array as the latitude is adjusted away from 0° (Fig. 4A). The nuclear power system is predicted to require ~9.5 t; hence, the PV+E system out-performs this option across ~50% of the planet's surface (Fig. 4B).

In addition to predicting production capacity and carry-along mass, our model provides design rules for optimal solar cell design. Optimal absorber bandgaps for the PV array are strongly dependent on the location on the surface of Mars (Fig. 4C-E). Several factors cause this variation: the total depth of the air column above a given location (*i.e.*, the difference between the height of the atmosphere and the altitude), gradients in dust and ice concentrations and particle radii, and orbital geometry effects that cause different effective air column thicknesses for locations near the poles. Lower elevations, higher dust and/or ice concentrations, and increasing distance away from the equator (near-polar latitudes) all cause

an increase in the optical depth of the air column, which enhances the fraction of light that is scattered. Because the spectrum of scattered light is slightly red-shifted with respect to direct light (Fig. 1F), optimal bandgaps decrease to capture more lower-energy photons (Fig. 4C-E) in regions where the optical depth is higher. For example, at equivalent latitudes, the optimal bandgaps are wider for regions with higher elevations than for those with lower elevations because the fraction of light that gets scattered is lower. Regional differences in atmospheric conditions can drive countervailing effects; because the Northern Hemisphere experiences generally lower dust concentrations than the Southern Hemisphere, the lower elevation in the Northern Hemisphere does not result in (on average) narrower optimal bandgaps. Instead, the reduced dust concentration (relative to that of the Southern Hemisphere) results in a reduced optical depth, resulting in wider optimal bandgap combinations (Fig. 4C-E).

In summary, solar cell arrays with careful attention to semiconductor choice and device construction represent a promising technology for sustaining an Earth-independent crewed habitat on Mars. Our analysis provides design rules for solar cells on the Martian surface and shows that solar cells can offer substantial reduction in carry-along mass requirements compared to alternative technology over a large fraction of the planet's surface.

**References**
1. J. T. Wilson, V. R. Eke, R. J. Massey, R. C. Elphic, W. C. Feldman, S. Maurice, L. F. A. Teodoro, Equatorial locations of water on Mars: Improved resolution maps based on Mars Odyssey Neutron Spectrometer data. *Icarus*. **299**, 148–160 (2018).
2. C. R. Webster, P. R. Mahaffy, G. J. Flesch, P. B. Niles, J. H. Jones, L. A. Leshin, S. K. Atreya, J. C. Stern, L. E. Christensen, T. Owen, H. Franz, R. O. Pepin, A. Steele, Isotope ratios of H, C, and O in CO2 and H2O of the martian atmosphere. *Science (80-. )*. **341**, 260–263 (2013).
3. G. Marbán, T. Valdés-Solís, Towards the hydrogen economy? *Int. J. Hydrogen Energy*. **32**, 1625–1637 (2007).
4. A. J. Berliner, J. M. Hilzinger, A. J. Abel, M. J. McNulty, G. Makrygiorgos, N. J. H. Averesch, S. Sen Gupta, A. Benvenuti, D. F. Caddell, S. Cestellos-Blanco, A. Doloman, S. Friedline, D. Ho, W. Gu, A. Hill, P. Kusuma, I. Lipsky, M. Mirkovic, J. Luis Meraz, V. Pane, K. B. Sander, F. Shi, J. M. Skerker, A. Styer, K. Valgardson, K. Wetmore, S.-G. Woo, Y. Xiong, K. Yates, C. Zhang, S. Zhen, B. Bugbee, D. S. Clark, D. Coleman-Derr, A. Mesbah, S. Nandi, R. M. Waymouth, P. Yang, C. S. Criddle, K. A. McDonald, L. C. Seefeldt, A. A. Menezes, A. P. Arkin, Towards a Biomanufactory on Mars. *Front. Astron. Sp. Sci.* **8**, 1–14 (2021).
5. J. G. Chen, R. M. Crooks, L. C. Seefeldt, K. L. Bren, R. M. Bullock, M. Y. Darensbourg, P. L. Holland, B. Hoffman, M. J. Janik, A. K. Jones, M. G. Kanatzidis, P. King, K. M. Lancaster, S. V Lymar, P. Pfromm, W. F. Schneider, R. R. Schrock, Beyond fossil fuel–driven nitrogen transformations. *Science (80-. )*. **360**, eaar6611 (2018).
6. A. Ursúa, L. M. Gandía, P. Sanchis, Hydrogen production from water electrolysis: Current status and future trends. *Proc. IEEE*. **100**, 410–426 (2012).
7. W. H. Cheng, M. H. Richter, M. M. May, J. Ohlmann, D. Lackner, F. Dimroth, T. Hannappel, H. A. Atwater, H. J. Lewerenz, Monolithic Photoelectrochemical Device for Direct Water Splitting with 19% Efficiency. *ACS Energy Lett.* **3**, 1795–1800 (2018).
8. A. A. Menezes, J. Cumbers, J. A. Hogan, A. P. Arkin, Towards synthetic biological approaches to resource utilization on space missions. *J. R. Soc. Interface*. **12** (2015), doi:10.1098/rsif.2014.0715.
9. S. N. Nangle, M. Y. Wolfson, L. Hartsough, N. J. Ma, C. E. Mason, M. Merighi, V. Nathan, P. A. Silver, M. Simon, J. Swett, D. B. Thompson, M. Ziesack, The case for biotech on Mars. *Nat. Biotechnol.* **38**, 401–407 (2020).
10. B. G. Drake, S. J. Hoffman, D. W. Beaty, Human exploration of Mars, design reference

architecture 5.0. *IEEE Aerosp. Conf. Proc.*, 1–24 (2010).
11. P. L. Read, M. Collins, F. Forget, R. Fournier, F. Hourdin, S. R. Lewis, O. Talagrand, F. W. Taylor, N. P. J. Thomas, A GCM climate database for Mars: For mission planning and for scientific studies. *Adv. Sp. Res.* **19**, 1213–1222 (1997).
12. B. Mayer, A. Kylling, Technical note: The libRadtran software package for radiative transfer calculations -- description and examples of use. *Atmos. Chem. Phys. Discuss.* **5**, 1319–1381 (2005).
13. Á. Vicente-Retortillo, F. Valero, L. Vázquez, G. M. Martínez, A model to calculate solar radiation fluxes on the Martian surface. *J. Sp. Weather Sp. Clim.* **5**, A33 (2015).
14. M. C. Hanna, A. J. Nozik, Solar conversion efficiency of photovoltaic and photoelectrolysis cells with carrier multiplication absorbers. *J. Appl. Phys.* **100** (2006), doi:10.1063/1.2356795.
15. H. Döscher, J. F. Geisz, T. G. Deutsch, J. A. Turner, Sunlight absorption in water – efficiency and design implications for photoelectrochemical devices. *Energy Environ. Sci.* **7**, 2951–2956 (2014).
16. G. A. Landis, in *Conference Record of the Thirty-first IEEE Photovoltaic Specialists Conference, 2005.* (IEEE; http://ieeexplore.ieee.org/document/1488268/), pp. 858–861.
17. P. M. Stella, R. C. Ewell, J. J. Hoskin, in *Conference Record of the Thirty-first IEEE Photovoltaic Specialists Conference, 2005.* (IEEE; http://ieeexplore.ieee.org/document/1488209/), pp. 626–630.
18. J. L. Young, M. A. Steiner, H. Döscher, R. M. France, J. A. Turner, T. G. Deutsch, Direct solar-to-hydrogen conversion via inverted metamorphic multi-junction semiconductor architectures. *Nat. Energy*. **2**, 17028 (2017).
19. J. Jia, L. C. Seitz, J. D. Benck, Y. Huo, Y. Chen, J. W. D. Ng, T. Bilir, J. S. Harris, T. F. Jaramillo, Solar water splitting by photovoltaic-electrolysis with a solar-to-hydrogen efficiency over 30%. *Nat. Commun.* **7**, 1–6 (2016).
20. S. Hu, C. Xiang, S. Haussener, A. D. Berger, N. S. Lewis, An analysis of the optimal band gaps of light absorbers in integrated tandem photoelectrochemical water-splitting systems. *Energy Environ. Sci.* **6**, 2984–2993 (2013).

**Author contributions:**
Conceptualization: AJA, AJB
Data curation: AJB, AJA
Funding acquisition: AJB, AJA, APA, DSC
Methodology: AJA, AJB, WDC
Software: AJB, AJA, MM
Supervision: WDC, APA, DSC
Visualization: AJA, AJB
Writing – original draft: AJA, AJB

Writing – review & editing: AJA, AJB, MM, WDC, APA, DSC

**Competing interests:** The authors declare no competing interests.

**Data availability:** All redSun software is available through the CUBES github organization at https://github.com/cubes-space/redSun and all data is available upon request.

**Supplementary Materials**
Methods; Supplementary Text; Figs. S1 to S13; Tables S1 to S8; References (1-23)

**Figures**

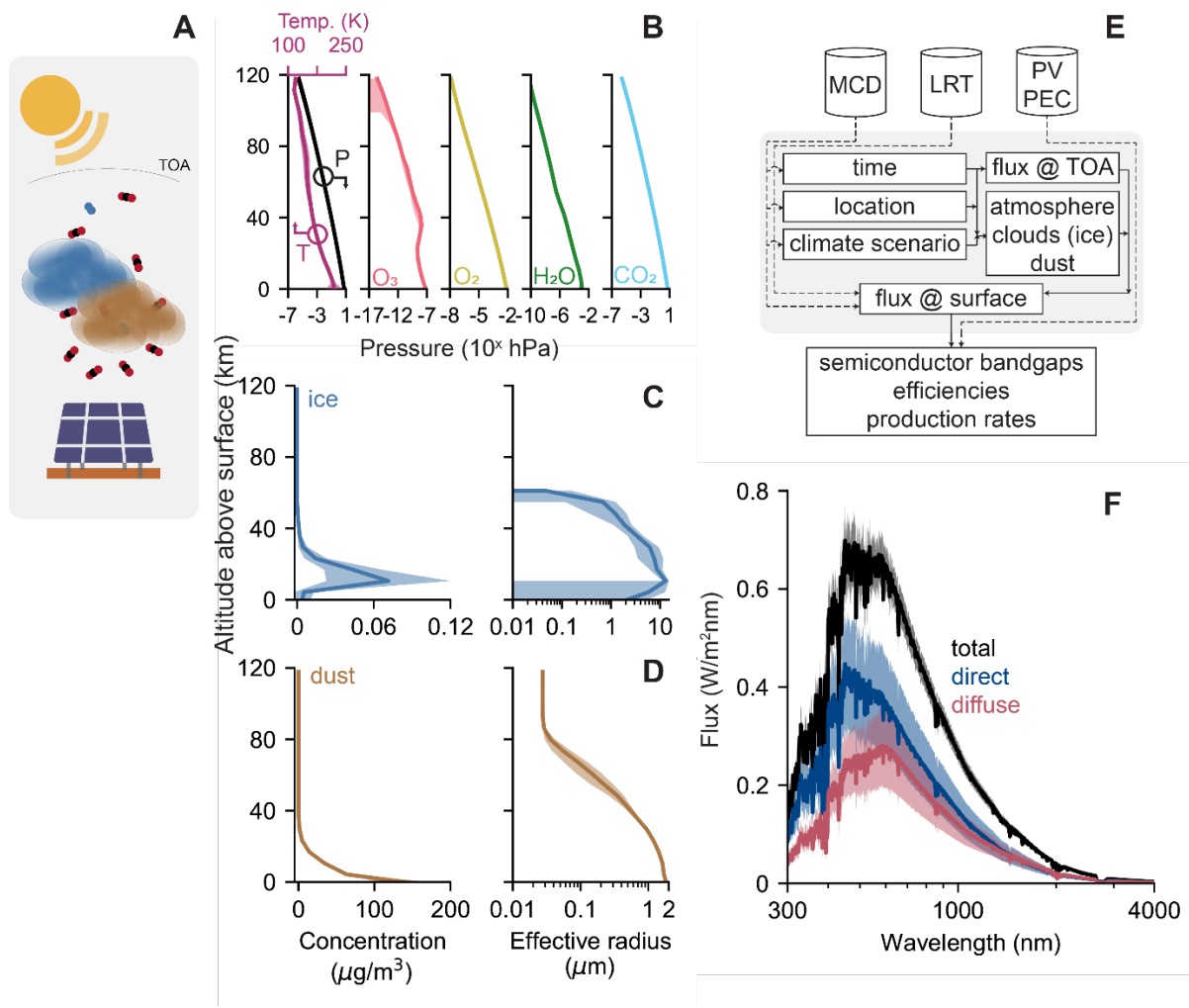

**Figure 1. Overview and calculation of spectral flux using atmospheric data.** (**A**) Sunlight incident on the solar cells is mediated by orbital geometry and local atmospheric composition of gases, ice, and dust. (**B**, **C**, and **D**) Temperature, partial pressure of atmospheric gases and concentration and effective radii of ice and dust particles as a function of altitude above the surface. (**E**) Information flow in the calculation scheme. Dotted lines represent functions used for calculations; solid lines represent data used as parameters. MCD, Mars Climate Database; LRT, LibRadtran. (**F**) Total (black), direct (blue), and diffuse (red) solar flux at Jezero Crater at solar noon averaged over the course of a typical Martian year. In (**B**),

(**C**), (**D**), and (**F**), solid lines represent yearly averages and shaded regions represent the standard deviation due to seasonal variation.

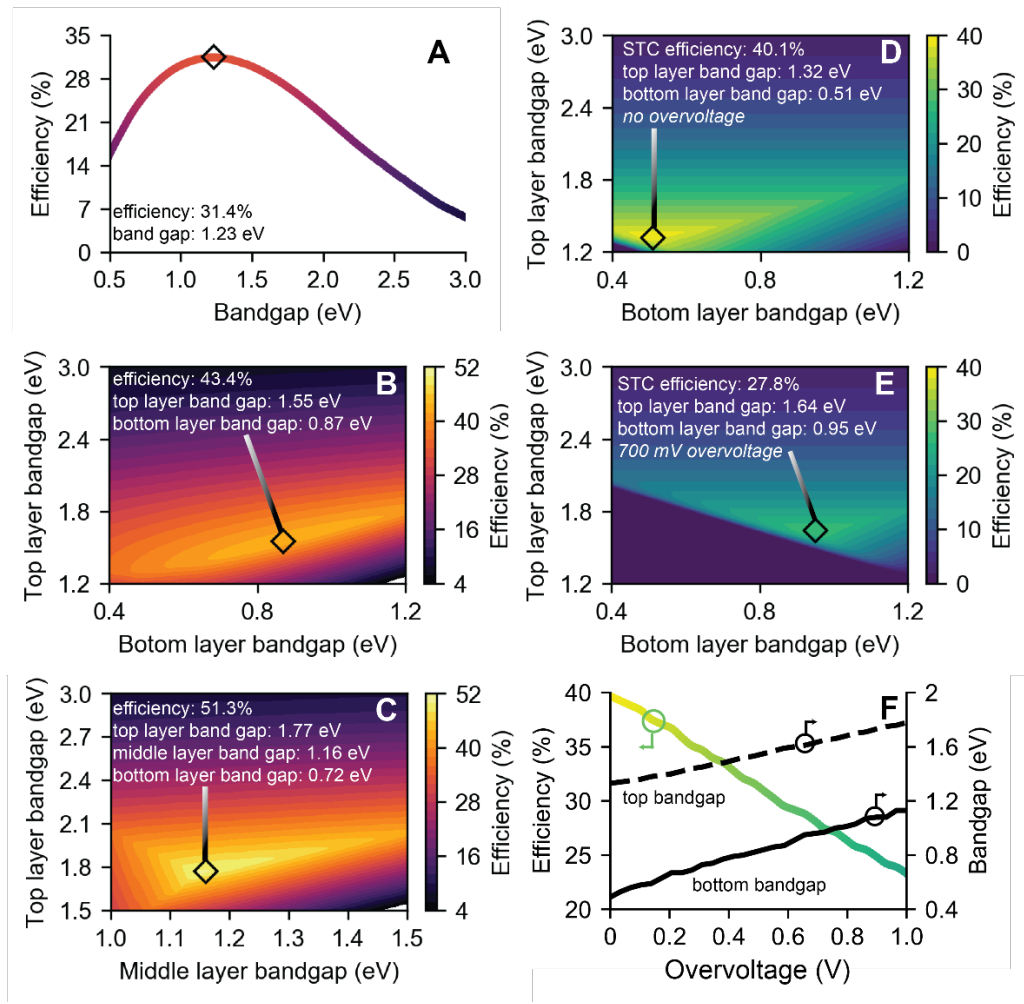

**Figure 2. Theoretical efficiencies of PV and PEC devices.** Detailed-balance efficiency limits as a function of bandgap energies for (**A**) single-junction, (**B**) two-junction, (**C**) three-junction photovoltaic devices. (**D**, and **E**) Solar-to-chemical (STC) efficiency for two-junction water splitting PEC devices producing molecular hydrogen with 0 mV (**D**) and 700 mV (**E**) overvoltage. (**F**) STC efficiency and optimal bandgaps for two-junction $H_2$-generating PEC devices as a function of overvoltage. Coloring in (**A**) and (**F**) correspond to contour coloring in (**B**, **C**) and (**D**, **E**) respectively. Average flux at solar noon at Jezero Crater is used as the reference solar spectrum.

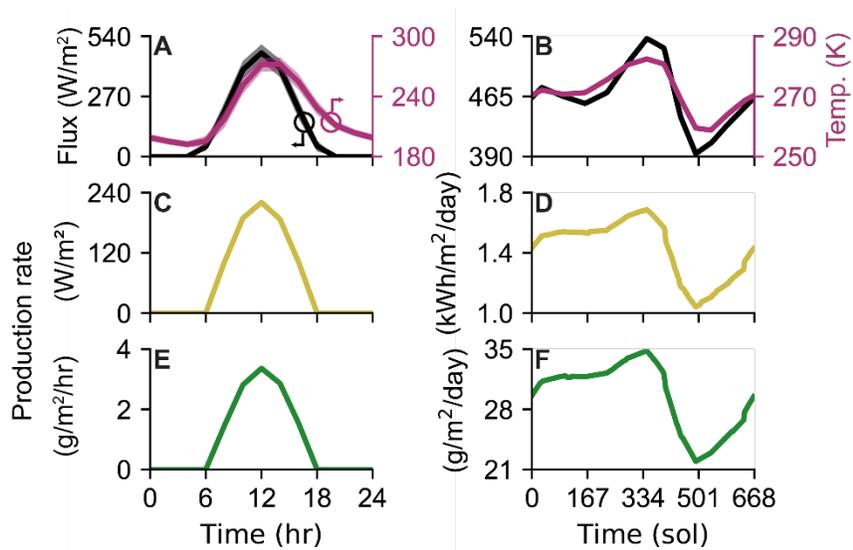

**Figure 3. PV and PEC production rates.** (**A**) Average and (**B**) daily maximum solar flux (black, left axis) and surface temperature (purple, right axis) as function of (**A**) time of day and (**B**) time of year. (**C**, **E**) average and (**D**, **F**) daily maximum production capacity of power (**C**, **D**) and $H_2$ (**E**, **F**) using 3-junction PV and 2-junction PEC cells as described in the main text. Solid lines in (**A**, **C**, **E**) correspond to averages; shaded areas represent the standard deviation due to seasonal variation. Jezero Crater is used as the location for plots.

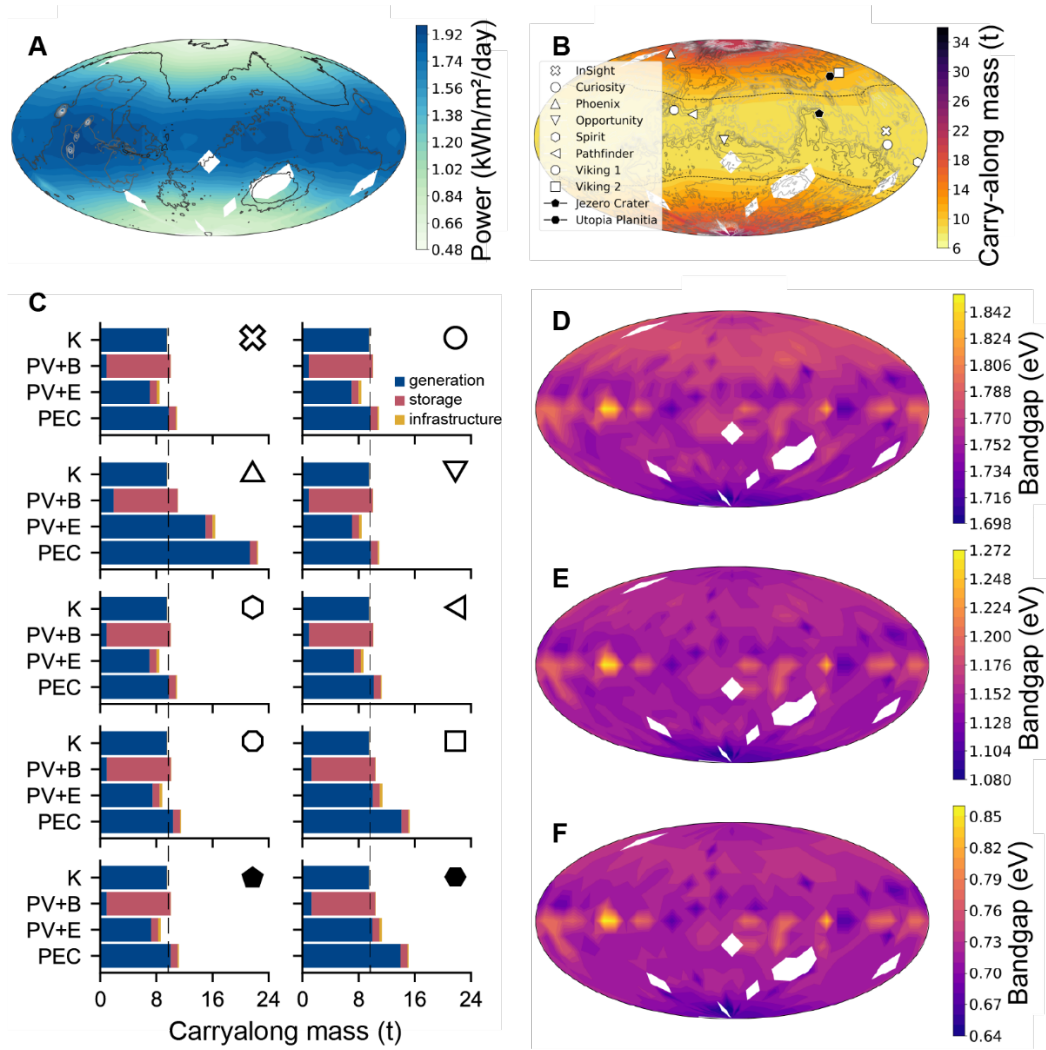

**Figure 4. Solar productivity across the Martian surface.** (**A**) Average daily solar power production capacity across the Martian surface. (**B**) Total carry-along mass required for power production using the PV+E generation system. Black dashed lines in (**A**) and (**B**) correspond to breakeven mass with nuclear power generation. Optimal (**C**) top, (**D**) middle, (**E**) bottom bandgaps for the 3-junction PV array.



\maketitle

\thispagestyle{empty}

\section{Introduction and Overview}


\begin{figure}[h!]
\centering
\includegraphics[width=0.8\textwidth]{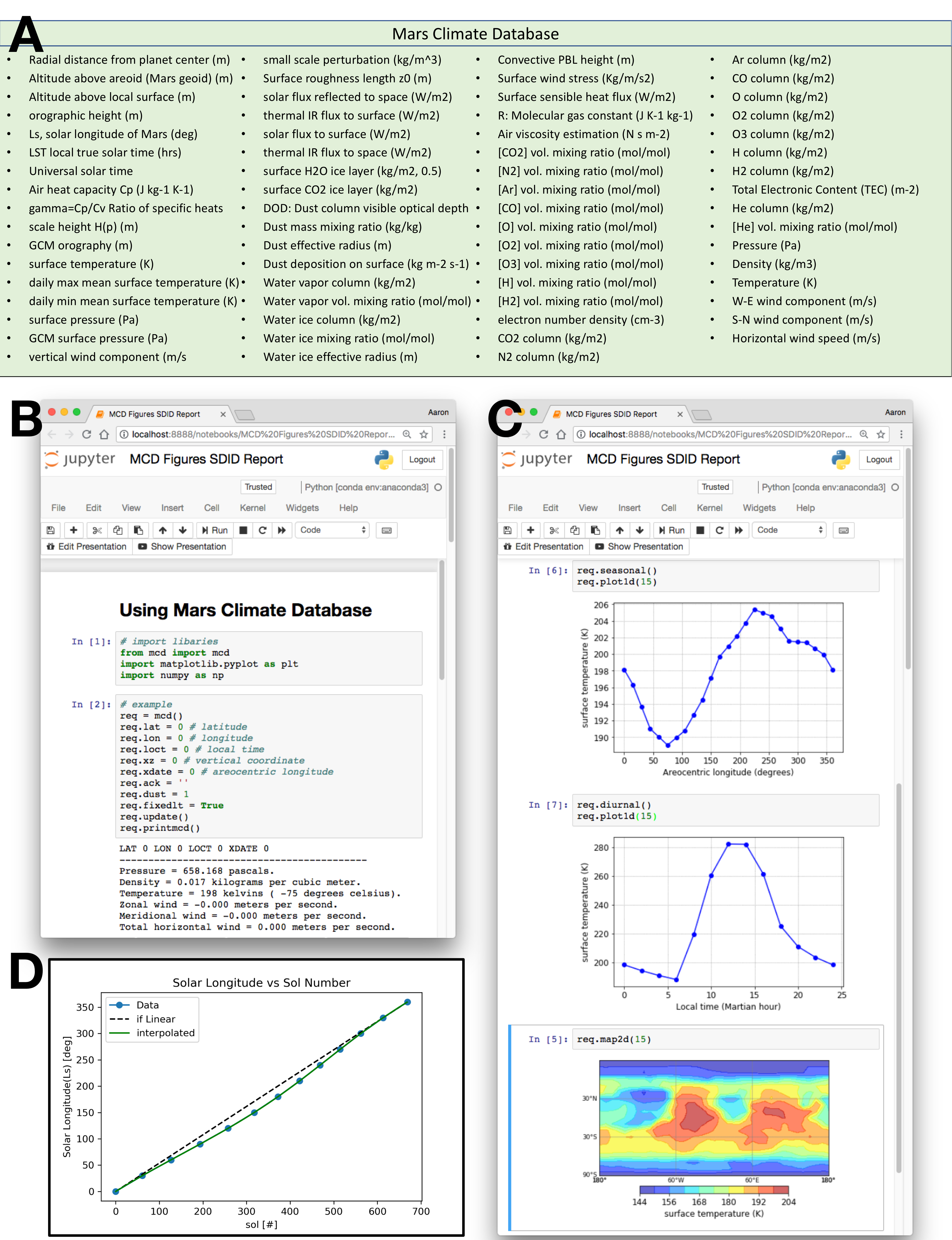}
\caption{\textbf{(A)} all parameters available for query in the MCD; \textbf{(B.)} example query to MCD; \textbf{(C.)} i. plot of surface temperature vs areocentric longitude for local time t = 9:00; ii. plot of surface temperature vs local time for LATITUDE = LONGITUDE = 0; iii. cylindrical projection of surface temperature; \textbf{(D.)} Plot of solar longitude vs sol number, demonstrating the eccentricity of Mars's orbit and the approximate season, with northern summer solstice occurring when Ls = 90 and northern winter solstice when Ls = 270.}
\label{MCD-1}
\end{figure}

A central question surrounding possible settlement of Mars is whether human life can be supported by available technologies using \textit{in situ} resources. Here we present a detailed analysis showing that photovoltaic and photoelectrochemical devices would be adequate and practical to sustain a crewed outpost for an extended period over a large fraction of the planet’s surface. Climate data were integrated with a radiative transfer model to predict spectrally-resolved solar flux across the Martian surface, which informed detailed balance calculations for solar cell devices supporting power systems, agriculture, and manufacturing. Optimal design and the corresponding production capacity over a Martian year revealed the size and mass of a solar cell array required to support a six-person mission, which represents less than 10\% of the anticipated payload.  

The following SI describes the \lstinline{redSun} software created as an integration of available software and custom code written in Python 3.6 with UNIX and Fortran backends. It can be found at \url{https://github.com/cubes-space/redSun}.

\section{Environmental Data Aggregation}

\subsection{Mars Climate Database}

Downstream radiative transfer calculations require a number of input streams describing the Martian environment. We make use of the Mars Climate Database (MCD) \cite{Bingham2003} developed by Le Laboratoire de Meteorologie Dynamique (LMD) in Paris, queried via the mcd-python package, to model most climate and environmental constraints, including photon flux and power spectra over time and location. The software engineering processes for building and using MCD somewhat efficiently are illustrated in Figure \ref{MCD-1}, along with input parameter profiles and sample output plots.


\subsection{Initial Geotemporalspatial Grid}
We began by first initializing the geotemporalspatial grid from which all downstream radiative transfer and PV/PEC calculations would be based. The grid was composed as a \lstinline{.netCDF} file with dimensions of 19 points of 10$^{\circ}$ latitude $\times$ 37 points of 10$^{\circ}$ longitude $\times$ 25 points of 15$^{\circ}$ areocentric longitude $\times$ 13 points of 2 (Martian) hours. Additionally, we included the dimension of altitude above the Martian datum in 20 points ranging from 0 to 120 km. The dimensions for the initial grid are shown in Table \ref{tab:igrid-dim}. 
\begin{table}[h!]
\centering
\begin{tabular}{@{}llllll@{}}
\toprule
Dimension             & Units         & Initial & Final & Step & Number \\ \midrule
Latitude              & degrees north & -90     & 90    & 15   & 19     \\
Longitude             & degrees east  & -180    & 180   & 15   & 37     \\
Wavelength            & nm            & 300.5   & 4000  & N/A  & 1340   \\
Level                 & km            & 0       & 120   & 6.32 & 20     \\
Aerocentric Longitude & deg           & 0       & 360   & 15   & 25     \\
Hour                  & hr            & 0       & 24    & 2    & 13     \\ \bottomrule
\end{tabular}
\caption{Initial grid dimensions.}
\label{tab:igrid-dim}
\end{table}

\subsection{Atmospheric Variables}
Through a combination of custom code in redSun and modifications to the Python-based extension of MCD, we then looped through Lat, Lon, Hr, and Ls dimensions to initialize the data variables in Table \ref{tab:igrid-var}.

\begin{table}[h!]
\centering
\begin{tabular}{@{}llll@{}}
\toprule
Variable                     & Units                       & Dimensions          & Dimension Number \\ \midrule
Air Density                  & cm$^{-3}$                        & lat,lon,level,ls,hr & 5                \\
Datum Altitude               & km                          & lat,lon,level       & 3                \\
CO$_2$ Partial Pressure         & cm$^{-3}$                       & lat,lon,level,ls,hr & 5                \\
H$_{2}$O Partial Pressure         & cm$^{-3}$                       & lat,lon,level,ls,hr & 5                \\
O$_{2}$ Partial Pressure          & cm$^{-3}$                       & lat,lon,level,ls,hr & 5                \\
O$_{3}$ Partial Pressure          & cm$^{-3}$                        & lat,lon,level,ls,hr & 5                \\
NO$_{2}$ Partial Pressure         & cm$^{-3}$                       & lat,lon,level,ls,hr & 5                \\
Pressure                     & hPa                         & lat,lon,level,ls,hr & 5                \\
Temperature                  & K                           & lat,lon,level,ls,hr & 5                \\
Ice Content                  & g/m$^{3}$                        & lat,lon,level,ls,hr & 5                \\
Ice Effective Radius         & um                          & lat,lon,level,ls,hr & 5                \\
Dust Content                 & g/m$^{3}$                        & lat,lon,level,ls,hr & 5                \\
Dust Effective Radius        & $\mu$m                          & lat,lon,level,ls,hr & 5                \\
Long Wave Downward Flux      & W/m$^{2}$      & lat,lon,ls,hr       & 4                \\
Short Wave Downward Flux     & W/m$^{2}$      & lat,lon,ls,hr       & 4                \\
Long Wave Upward Flux        & W/m$^{2}$     & lat,lon,ls,hr       & 4                \\
Short Wave Upward Flux       & W/m$^{2}$     & lat,lon,ls,hr       & 4                \\
Top of Atmosphere Irradiance & W/(nm.m$^{2}$) & lat,ls,hr,wl        & 5                \\ \bottomrule
\end{tabular}
\caption{Initial atmospheric grid variables sourced from MCD.}
\label{tab:igrid-var}
\end{table}

\subsection{Planetary Variables}
\begin{figure}[h!]
\centering
\includegraphics[width=.99\textwidth]{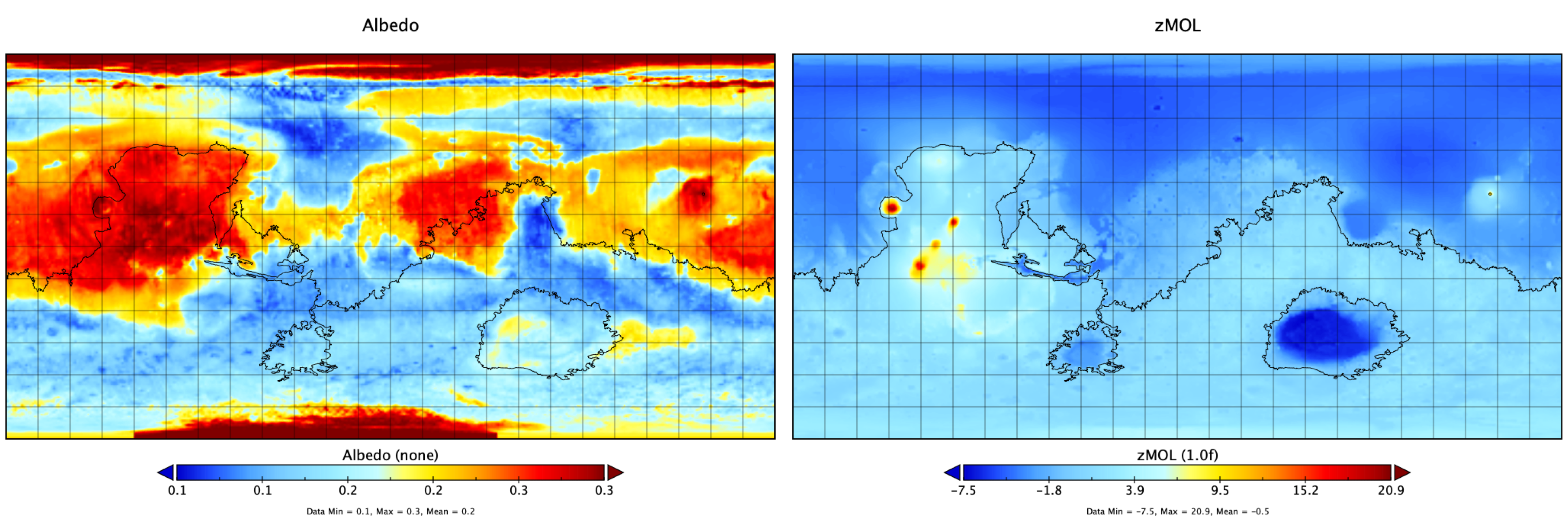}
\caption{Albedo and zMOL (height above the Martian datum) maps.}
\label{albedo}
\end{figure}

While most of the required environmental variables could be sourced from MCD, additional efforts were made to add data on the planetary albedo and zMOL as shown in Figure \ref{albedo} and in Table \ref{tab:igrid-pvar}.

\begin{table}[h!]
\centering
\begin{tabular}{@{}llll@{}}
\toprule
Variable & Units & Dimensions & Dimension Number \\ \midrule
Albedo   & None  & lat,lon    & 2                \\
zMOL     & None  & lat,lon    & 2                \\ \bottomrule
\end{tabular}
\caption{Initial planetary grid variables sourced from MCD.}
\label{tab:igrid-pvar}
\end{table}

\subsection{Solar Variables}
In addition to atmospheric and planetary variables, our initial environmental data for downstream radioactive transfer required that we calculate the solar flux at the top of the atmosphere (TOA). Downstream radiative transfer calculations required as input the spectral flux in W/(m$^{2}\cdot$nm) whereas MCD only provided an integrated solar flux in W/(m$^{2}$). For a given Lat, Lon, Hr, and Ls, we were able to calculate the spectral flux $F_{0}$ via\cite{Patel2004AnnualSurface}
\begin{align}
    F_{0} &= \mu F_{1.52}\left(\frac{d^{2}}{r^{2}}\right)\\
    F_{0} &= F_{1.52}\left(\sin\theta\sin\epsilon\sin L_{s} + \cos\theta\cos\left(\frac{2\pi t}{P}\right)\left(1-\sin^{2}\epsilon\sin^{2}L_{s}\right)^{1/2}\right)\left(\frac{1+e\cos(L_{s}-L_{s,p})}{1-e^{2}}\right)^{2}
\end{align}
where $r$ is the Sun-Mars distance along its orbit, $d$ is the mean Sun-Mars distance of 1.52 AU, $\mu$ is the cosine of the solar zenith angle $z$, $e$ is the Martian eccentricity ($e=0.0934)$, $L_{s}$ is the aerocentric longitude,  $L_{s,p}$ is the aerocentric longitude of perihelion (250$^{\circ}$), $\theta$ is the latitude, $\epsilon$ is the Martian obliquity (25.2$^{\circ}$), $P$ is the duration of the Martian solar day (88775 s), $t$ is any time measured from local noon, and $F_{1.52}$ is the flux at the average Sun-Mars distance\cite{Vicente-Retortillo2015ASurface}. 

While the separation of the aerocentric longitude and hourly time dimensions was helpful in indexing our grid, these two dimensions are related. For any aerocentric longitude index, there are 13 time points, and as these times correspond to movement of Mars around the sun, so does the aerocentric longitude. Therefore, when computing the TOA flux $F_{0}$, we updated $L_{s}$ to correspond to the change in time $t$ using the build in functions \lstinline{Ls2Sol} and \lstinline{Sol2Ls} from the \lstinline{MCD} package. These functions relate $L_{s}$ and $t$ through Kepler's Problem via
\begin{align}
    L_{s} &= \left(\nu \frac{180}{\pi} + L_{s,p}\right)(mod360)\\
    \nu &= 2\arctan \left[ \sqrt{\frac{1+e}{1-e}}\tan\left(\frac{E}{2}\right)\right]\\
    M &= E - e\sin E = 2\pi\frac{D_{s}-t_{p}}{N_{s}}
\end{align}
where $D_{s}$ is the sol number, $ t_{peri}$ is the time at perihelion, $N_{s}$ is the number of sols in a Martian year, $\nu$ is the true anomaly, $E$ is the eccentric anomaly, $M$ is the mean anomaly, and $N_{s}$ is the number of sols in a Martian year. 

\begin{figure}[h!]
\centering
\includegraphics[width=.99\textwidth]{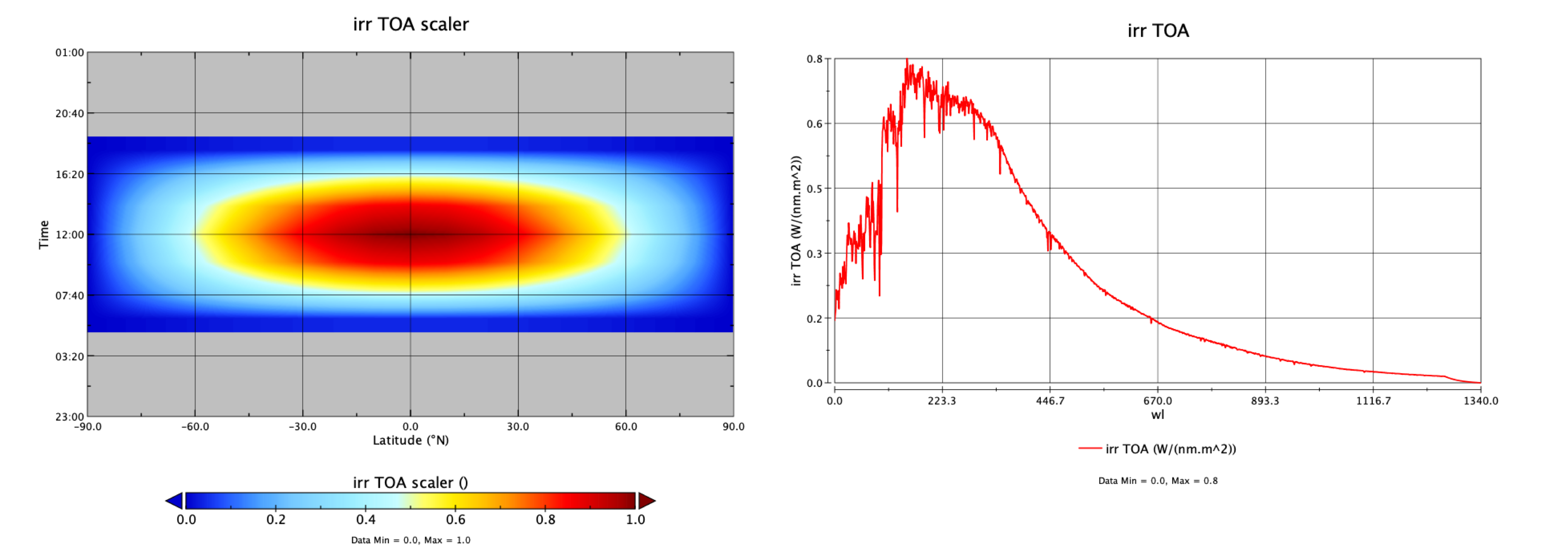}
\caption{Left shows the calculated $mu$ parameter as a scalar across geospace for $L_{s}=0$. Right shows the spectral flux for lat=0, $t$=12 noon, and $L_{s}=0$.}
\label{fluxtoa}
\end{figure}

The data variables shown in Figure \ref{fluxtoa} were then added to the grid for downstream use as shown in Table \ref{tab:igrid-fvar}.

\begin{table}[h!]
\centering
\begin{tabular}{@{}llll@{}}
\toprule
Variable                     & Units                       & Dimensions   & Dimension Number \\ \midrule
Solar Zenith Angle           & deg                         & lat,ls,hr    & 3                \\
Solar Correction             & None                        & lat,ls,hr    & 3                \\
Top of Atmosphere Irradiance & W/(nm.m$^{2}$) & lat,ls,hr,wl & 5                \\ \bottomrule
\end{tabular}
\caption{Initial solar grid variables.}
\label{tab:igrid-fvar}
\end{table}

\begin{figure}[h!]
\centering
\includegraphics[width=.8\textwidth]{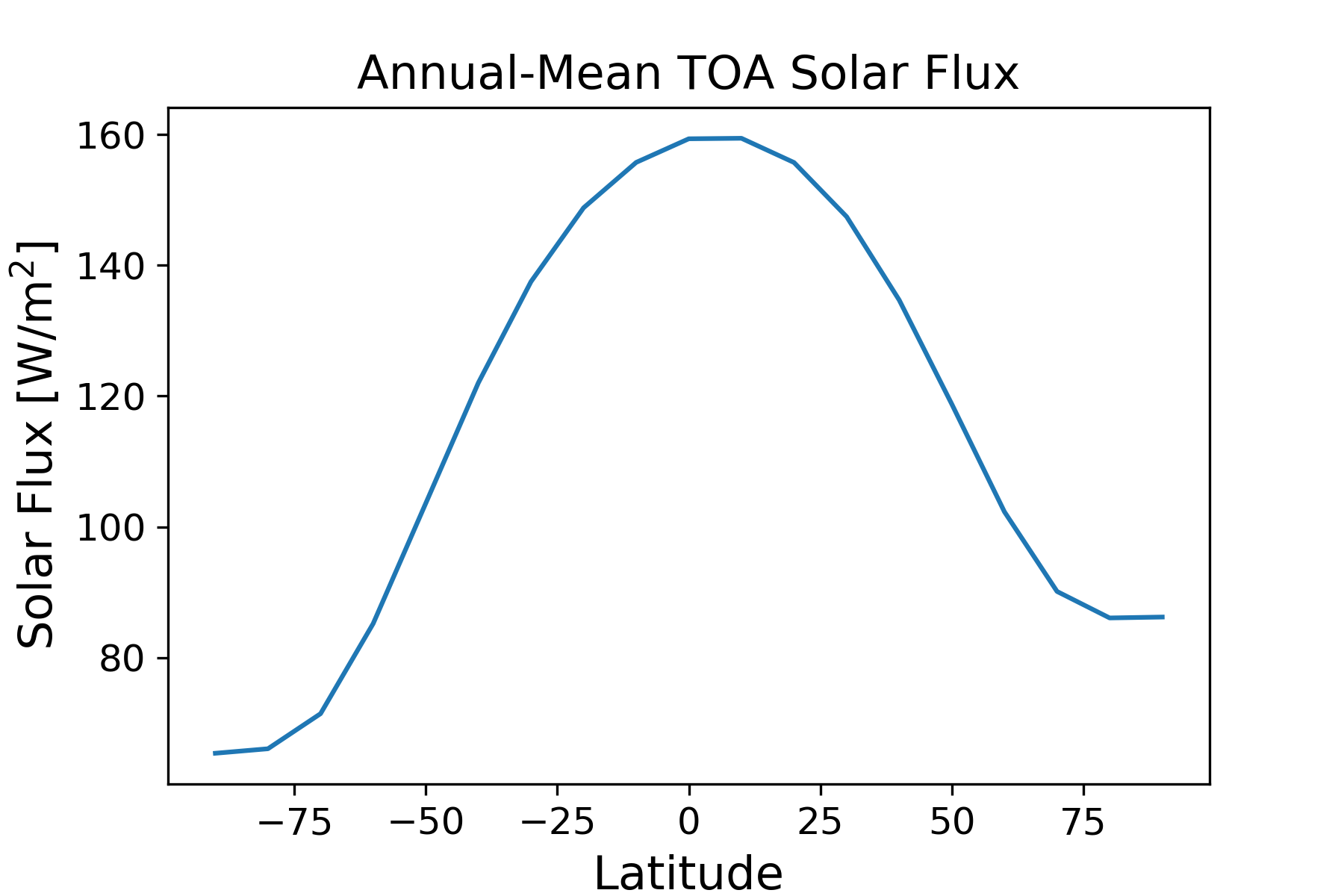}
\caption{Calculated Annual-Mean TOA Solar Flux distributed across Latitude}
\label{sanityflux}
\end{figure}

As a sanity check, we calculated the integrated standard solar flux at TOA at 1.52 AU (average Sun-Mars distance) at 576.92 W/m$^{2}$. Given a solar constant for Mars is 490 W/m$^2$, the equatorial annual-mean flux at the top of the atmosphere (TOA) should be $\sim$156 W/m$^2$. Our calculated equatorial annual-mean TOA flux was found to be 159.43 W/m$^2$ which differs by $\sim$1.5\% from the theoretical value. We extended this calculation across all latitudes as shown in Figure \ref{sanityflux} to confirm our methods.

\section{Radiative Transfer Calculations}

\subsection{libRadtran}
The radiative transfer calculations were carried out using the libRadtran library (version 2.0.4)\cite{Mayer2005,Emde2016The1}. libRadtran is a collection of C and Fortran functions and programs for calculation of solar and thermal radiation in the Earth's atmosphere and is freely available under the GNU General Public License at \url{http://www.libradtran.org/doku.php}.

\subsection{Mie Scattering Calculations}
The presence of dust and cloud particles in the Martian atmosphere affect the propagation of sunlight. The size of such dust and cloud particles falls within the Mie scattering range. 

\begin{figure}[h!]
\centering
\includegraphics[width=.8\textwidth]{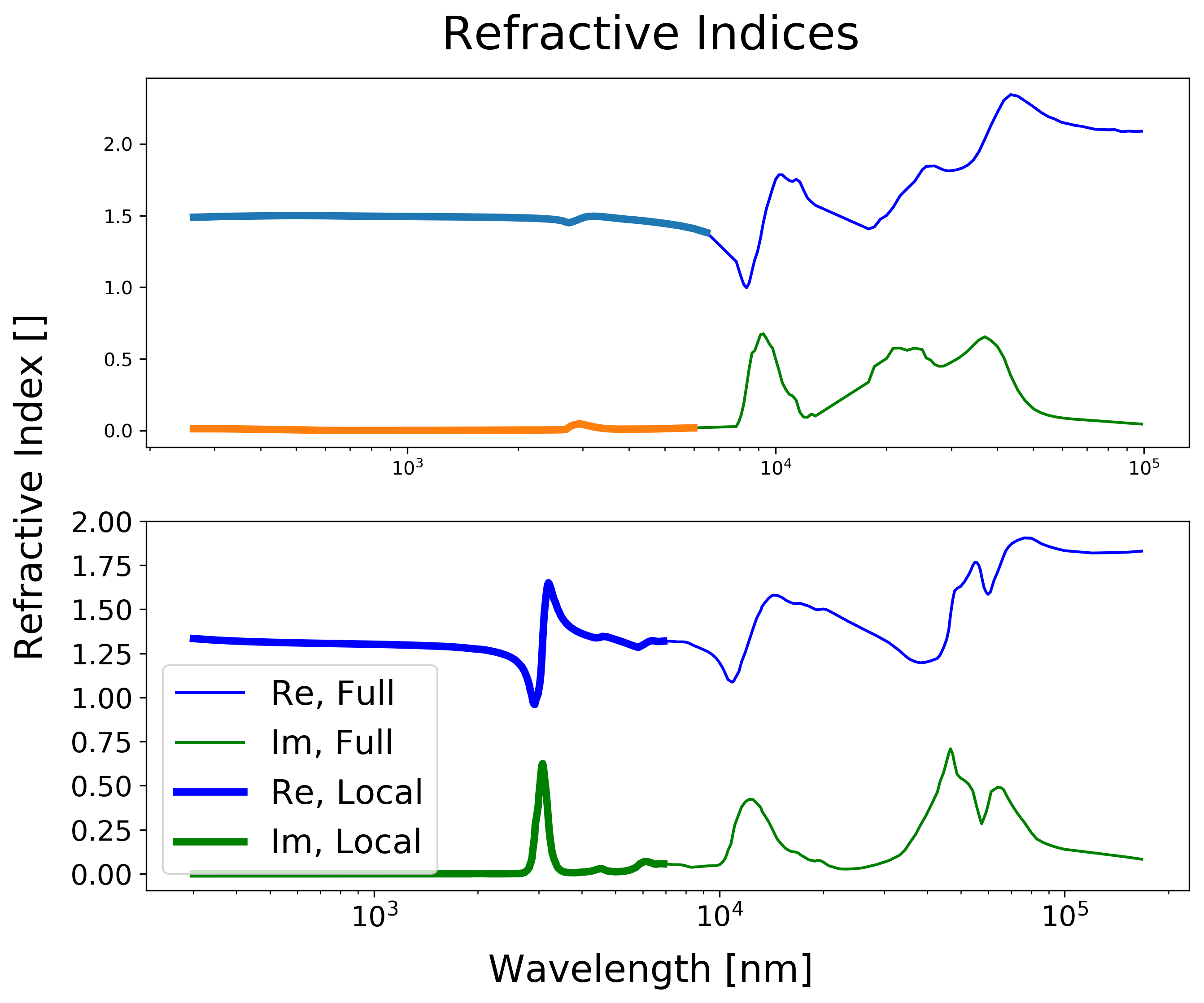}
\caption{Refractive Indices for Dust (top) and Clouds (Bottom).}
\label{refind}
\end{figure}

The libRadtran package was used for Mie scattering calculations of the scattering phase matrices and corresponding Legendre polynomials\cite{Wiscombe1980ImprovedAlgorithms}. Input files for both dust and ice were constructed (Listing \ref{mieinput}) and fed to the \lstinline{MIEV0} tool.
\begin{lstlisting}[language=bash, caption=Input file for Mie calculations of cloud aerosols, label={mieinput}]
mie_program MIEV0 # Select Mie code by Wiscombe
basename cloud.
refrac file MieCloudRefrac.DAT# Use refractive index file
r_eff 0.00322766 100.1 10.0  # Specify effective radius grid
distribution lognormal 1.8903 # Specify lognormal size distribution
nstokes 1 # Calculate all phase matrix elements
nmom 6000 # Number of Legendre terms to be computed
nthetamax 2000 # Maximum number of scattering angles to be
output_user netcdf # Write output to netcdf file
verbose # Print verbose output
\end{lstlisting}
Refractive indices for dust and ice were sourced from NASA Ames Legacy Mars Global Climate Model\cite{Haberle2019DocumentationCycle} (available at \url{https://github.com/nasa/legacy-mars-global-climate-model}) and fed as input (Figure \ref{refind}).

For clouds, an effective radius $r_{\text{eff}}$ grid was set between 0.00322766 and 100.1 $\mu$m in steps of 10 $\mu$m and with a lognormal distribution with standard deviation $\sigma=1.8903$ described as
\begin{align}
    n(r) & = \frac{a}{r}\exp\left(-\frac{1}{2}\left(\frac{\ln(r)\ln(r_{0})}{\ln\sigma}\right)^{2}\right)
\end{align}
where $r_{0}$ is the logarithmic mode of the distribution, calculated from $r_{\text{eff}}$. Through a series of trial-and-error attempts, we specified additional parameters for clouds such as the number of phase matrix elements set at 1, the number of Legendre terms to be computed set at 6000, the maximum number of scattering angles set to 2000. The resulting output from \lstinline{MIEV0} was a \lstinline{.netCDF} file of $\sim$100 MB.

For dust, an effective radius $r_{\text{eff}}$ grid was set between 0.00310352 and 10.1 $\mu$m in steps of 1.0 $\mu$m and with a lognormal distribution with standard deviation $\sigma=1.3616$. Again, through a series of trial-and-error attempts, we specified additional parameters for dust such as the number of phase matrix elements set at 1, the number of Legendre terms to be computed set at 2500, the maximum number of scattering angles set to 2000. The dust calculations provided more computationally expensive than those for clouds due to the smaller $r_{\text{eff}}$ grid size. The resulting output from \lstinline{MIEV0} was a \lstinline{.netCDF} file of $\sim$10 MB.

\begin{figure}[h!]
\centering
\includegraphics[width=1.0\textwidth]{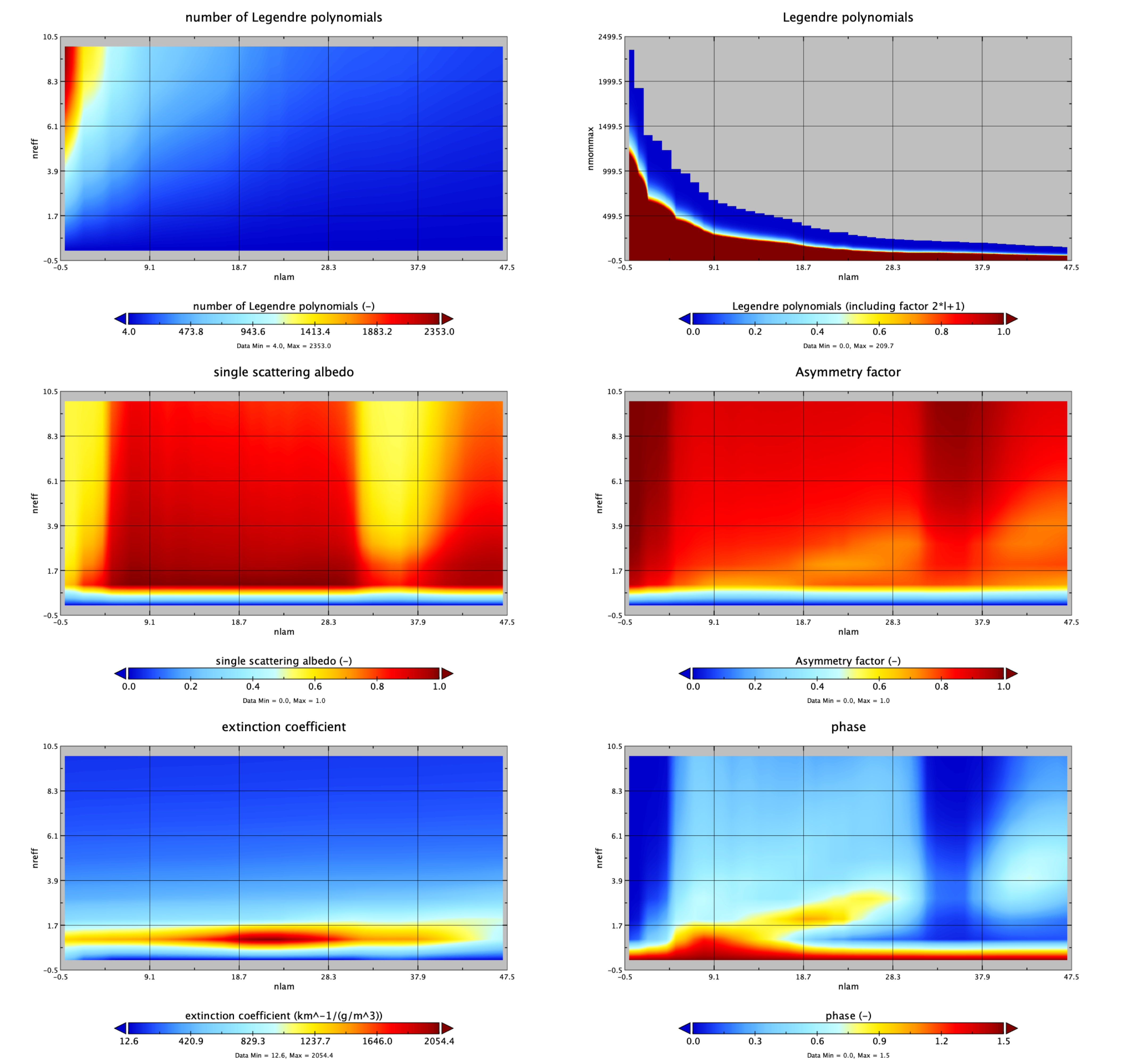}
\caption{Sample Visualization of variables in \lstinline{.netCDF} Mie output file for dust.}
\label{mievis}
\end{figure}

The output \lstinline{.netCDF} files include the dimensions and variables in Table \ref{tab:mievar} and a sample of the output variables are shown in Figure \ref{mievis}.

\begin{table}[h!]
\centering
\begin{tabular}{@{}llll@{}}
\toprule
Name      & Description                    & Dim/Var & Unit                                            \\ \midrule
nlam      & Wavelength Number              & Dim     & -                                               \\
nmommax   & Legendre Polynomial Number     & Dim     & -                                               \\
nphamat   & Phase Matrix Element Number    & Dim     & -                                               \\
nreff     & Refractive Index Number         & Dim     & -                                               \\
nthetamax & Theta Max Number               & Dim     & -                                               \\
nrho      & Density Number                 & Dim     & -                                               \\
wavelen   & Wavelength                     & Var     & micrometer                                      \\
reff      & Effective radius               & Var     & micrometer                                      \\
ntheta    & Number of scattering angles    & Var     & -                                               \\
theta     & Theta Max Number               & Var     & degrees                                         \\
phase     & phase                          & Var     & -                                               \\
nmom      & number of Legendre polynomials & Var     & -                                               \\
pmom      & Legendre polynomials           & Var     & including factor 2*l+1                          \\
ext       & extinction coefficient         & Var     & km\textasciicircum{}-1/(g/m\textasciicircum{}3) \\
ssa       & single scattering albedo       & Var     & -                                               \\
gg        & Asymmetry factor               & Var     & -                                               \\
refre     & refractive index (real)        & Var     & -                                               \\
refim     & refractive index (imaginary)   & Var     & -                                               \\
rho       & density of medium              & Var     & g/cm\textasciicircum{}3                         \\ \bottomrule
\end{tabular}
\caption{Dimensions and variables in \lstinline{.netCDF} Mie output file.}
\label{tab:mievar}
\end{table}

\cleardoublepage
\subsection{uvspec}
The \lstinline{uvspec} program was designed to calculate the radiation field of the atmosphere for Earth. Modifications were carried out such that \lstinline{uvspec} could be leveraged for similar calculations of the Martian radiative transfer. Input to the model are the constituents of the atmosphere including various molecules, aerosols and clouds. The absorption and scattering properties of these constituents were calculated via the \lstinline{MIEV0} tool. Boundary conditions are the solar spectrum at the top of the atmosphere and the reflecting surface at the bottom\cite{Mayer2012LibRadtranGuide}. The \lstinline{uvspec} program was called for each point in the geotemporalspatial grid and provided with a custom, programmatically generated input file -- an example of which is shown in Listing \ref{uvspecinput}.
\begin{lstlisting}[language=bash, caption=Sample input file for uvspec calculation, label={uvspecinput}]
# libRadtran Calc test
wavelength 300.5 4000 # choose wavelength range for computation
atmosphere_file __2WKSII17KGatmos.DAT # load atmosphere profile
mixing_ratio CH4 0.0 # update null mixing ratios
mixing_ratio N2O 0.0
mixing_ratio F11 0.0
mixing_ratio F12 0.0
mixing_ratio F22 0.0
altitude -0.48425 # specify altitude above datum
source solar __2WKSII17KGflux.DAT # load solar profile
# corrected for Sun-Mars Distance
# corrected for geometry
ic_file 1D __2WKSII17KGcloud.DAT # setup cloud profile (assuming water/ice clouds)
ic_properties MieCalc/cloud.mie.cdf interpolate
profile_file dust 1D __2WKSII17KGdust.DAT # setup dust profile (using aerosol type)
profile_properties dust MieCalc/dust.mie.cdf interpolate
earth_radius 3389.5 # reset earth_radius to Martian radius in [km]
rte_solver disort pseudospherical # choose radiative transfer solver
pseudospherical
number_of_streams 6 # choose number of streams
output_user lambda edir eglo edn eup enet esum # define output
albedo 0.3073502629995346 # choose albedo
\end{lstlisting}
Due to the peculiar way \lstinline{uvspec} must be called, input for atmosphere, solar flux, dust conditions, and cloud conditions are required in the form of text-based \lstinline{.DAT} files. Because multiple \lstinline{uvspec} calls were carried out in parallel, a random string was generated (``2WKSII17KG'' in the case of Listing \ref{uvspecinput}) and used to identify specific \lstinline{.DAT} files. For each point of the grid, an input \lstinline{.INP} file was created along with correspond \lstinline{.DAT} files for atmosphere, solar flux, dust conditions, and cloud conditions. The atmosphere file contained the altitude above sea level in km, pressure in hPa, temperature in K, air density in cm$^{-3}$, ozone density in cm$^{-3}$, Oxygen density in cm$^{-3}$, water vapor density in cm$^{-3}$, CO$_{2}$ density in cm$^{-3}$, and NO$_{2}$ density in cm$^{-3}$. The dust and cloud aerosol files contained altitude above sea level in km, dust/cloud content in kg/kg, and effective radius in $\mu$m. The solar flux file contains the wavelength in nm and the spectral flux for that wavelength in mW/(m$^{2}$nm). Data for each of these files was sourced from the MCD data organized in the \lstinline{Stupidgrid.nc} file and converted to the appropriate units using functions in the \lstinline{redSun} codebase.

The wavelength range was set from 300.5 to 4000 nm. This range was selected to match available data for solar flux and significance to downstream photovoltaic calculations. Wavelengths outside these bounds were found to have negligible impact on bandgap calculations or to require substantial computational efforts, and were thus ignored. The mixing ratios for atmospheric CH$_{4}$, N$_{2}$O, and greenhouse gases (GHG) F11, F12, and F22 were set to 0.0 to reflect the change from Earth to Mars conditions. The altitude for the location was also programmatically added to the input file to specify the exact position of the surface in relationship to the Martian datum. The filenames from the Mie scattering calculations for dust and cloud aerosols were passed as well. The radius of the planet was changed to the Martian value of 3389.5 km. The albedo of the grid-point was also provided programmatically. 

We selected the DIScrete ORdinate Radiative Transfer solvers (\lstinline{pseudospherical disort}) radiative transfer solver for our calculations using 6 streams. The discrete ordinate method was first developed in 1960 with significant updates in 1988 and 2000 and offer 1D calculations of radiance, irradiance, and actinic flux. We opted for pseudo-spherical methods to offset any spherical effects associated with using the smaller Martian geometry. In pseudo-spherical calculations, the monochromatic radiative transfer equation in 1D can be formulated as

{\scriptsize
\begin{align}
    \mu\frac{dI(\tau,\mu,\phi)}{\beta^{\text{ext}}d\tau} &= I(\tau,\mu,\phi) - \frac{\omega(r)}{4\pi}\int_{0}^{2\pi}d\phi'\int_{-1}^{1}d\mu'p(\tau,\mu,\phi; \mu',\phi')I(\tau,\mu',\phi')-(1-\omega(r))B[T(r)] - \frac{\omega(\tau)I^{0}}{4\pi}p(\tau,\mu,\phi,\mu_{0},\phi_{0})e^{f_{ch}(\tau,\mu_{0})}
\end{align}
}%
where $B[T(r)]$ is the Planck function, $\beta$ is an extinction coefficient, $\mu_{0}$ is the solar zenith angle, $\phi_{0}$ is the azimuth angle, $p$ is the phase function, and the single scattering albedo $\omega(r)$ is
\begin{align}
    \omega(r) &= \omega(r,\nu) = \frac{\beta^{\text{sca}}(r,\nu)}{\beta^{\text{ext}}(r,\nu)}
\end{align}
Additionally, $f_{ch}$ is the Chapman function\cite{Rees1989PhysicsAtmosphere,Dahlback1991ATwilight} for describing the extinction path in a spherical atmosphere and is formulated as
\begin{align}
    f_{ch}(r_{0},\mu_{0}) = \int_{r_{0}}^{\infty}\frac{\beta^{\text{ext}}(r,\nu)dr}{\sqrt{1-\left(\frac{R+r_{0}}{R+r}\right)^{2}(1-\mu_{0}^{2})}}
\end{align}
where $R$ is the planet radius and $r_{0}$ is the distance above the atmosphere. 

The output of each \lstinline{uvspec} call was a text-like file that was indexed with a matching random string identifier. Each file consisted of the direct, global, diffuse downward, diffuse upward, net and sum irradiance in mW/(m$^{2}$nm) for each nm in the input flux file. The output file was then read back with additional functions from \lstinline{redSun} for use in downstream calculations.

\section{Photovoltaic Power and Photoelectrochemical Commodity Calculations}

We use the detailed balance model to calculate the energy efficiency of one-, two-, and three-bandgap photovoltaic solar cells and one- and two-bandgap photoelectrochemical devices. This model has been used to calculate the limiting efficiency of ideal photovoltaic and photoelectrochemical devices for single and multiple bandgap architectures previously\cite{Hanna2006,Doscher2014SunlightDevices,Hu2013AnSystems}.

The current density (J)-voltage (V) dependence $J\left(V,E_{g}\right)$ for a single bandgap is given by
\begin{align}
    J\left(V,E_{g}\right) &= J_{G}(E_{g}) + J_{R}(V,E_{g}) 
\end{align}
where $J_G$ is the photogeneration current, $J_R$ is the recombination current due to radiative recombination, and $E_g$ is the bandgap of the absorber material. The generation current $J_{G}$ is calculated according to
\begin{align}
    J_{G}(E_{g}) &= q\int_{E_{g}}^{E_\text{max}}\Gamma(E)dE
\end{align}
where $q$ is the electronic charge, $\Gamma(E)$ is the photon flux at a given photon energy $E$, and $E_\text{max}$ is maximum photon energy in the solar spectrum. We used a minimum wavelength of 300 nm in our calculations, corresponding to a maximum photon energy of $\sim$4.14 eV because photons above 4 eV contribute negligibly to the photon flux\cite{Hanna2006}. The recombination current density $ J_{R}$ is calculated according to
\begin{align}
    J_{R}(V,E_{g}) &= \frac{2\pi q}{c^{2}h^{3}}\int_{E_{g}}^{\infty}\frac{E^{2}}{\exp\left(\frac{E-qV}{kT}\right)-1}dE
\end{align}
where $c$ is the speed of light in vacuum, $h$ is Planck’s constant, $k$ is Boltzmann’s constant, and $T$ is the temperature of the device (we assume the local surface temperature in these calculations).

The photovoltaic energy efficiency $\eta_{PV}$ at a given operating voltage is written as
\begin{align}
    \eta_{\text{PV}}(V,E_{g}) &= \frac{V}{F}J(V,E_{g})
\end{align}
where $F$ is the calculated total power flux at the Martian surface. The operating voltage can then be selected to maximize the efficiency for a given bandgap. In technoeconomic calculations (see below), we assume the device efficiency is 80\% of the calculated detailed balance limit to account for absorber material and device inefficiencies (i.e., nonradiative recombination losses not captured by the detailed balance limit).

The photoelectrochemical device energy efficiency $\eta_{\text{PEC}}$ is given by
\begin{align}
    \eta_{\text{PEC}}(V,E_{g}) &= \frac{E^{0}}{F}J(V,E_{g})
\end{align}
where $E^0$ is the minimum thermodynamic potential required to drive the electrochemical reaction (1.23 V for H$_{2}$ generation from water splitting). In practical devices, the operating voltage of the photoelectrochemical device will be larger than $E^0$ to account for anode and cathode overpotentials and resistive potential drop in the electrolyte and electrodes. Hence, for these devices the operating voltage is
\begin{align}
    V = E^0 + V_{o}
\end{align}
where $V_o$ is the overpotential associated with the above-mentioned losses. In all technoeconomic calculations (see below) we assume the overvoltage is 700 mV, corresponding to a practical minimum that also accounts for absorber material inefficiencies (\textit{i.e.}, nonradiative recombination losses not captured by the detailed balance limit)\cite{Doscher2014SunlightDevices}.

For two- and three-bandgap tandem devices, we assume the absorber layers are connected optically and electronically in series. Generation and recombination currents are calculated as described above, with the modification that $E_{\text{max}}$ is substituted with $E_{g,n-1}$ for absorber $n$ (counted sequentially starting with the top absorber) to reflect the assumption that each absorber layer is optically thick (i.e, absorbs all the above-bandgap light incident on its surface). In tandem devices, the total current density must be equal in each absorber layer, while the total operating voltage is given by the sum of the voltages developed across each cell. For example, for a three-absorber photovoltaic device
\begin{align}
    J(V) &= J_{1}(V_{1},E_{g,1}) = J_{2}(V_{2},E_{g,2}) = J_{3}(V_{3},E_{g,3})\\
    V &= V_{1} + V_{2} + V_{3}
\end{align}
For tandem devices, the efficiency is calculated analogously to the single-junction devices but as a function of each absorber bandgap.

\section{Grid Calculations via Parallel Computing}
\subsection{SinglePoint Calculation}
\begin{figure}[h!]
    \centering
    \includegraphics[width=0.95\textwidth]{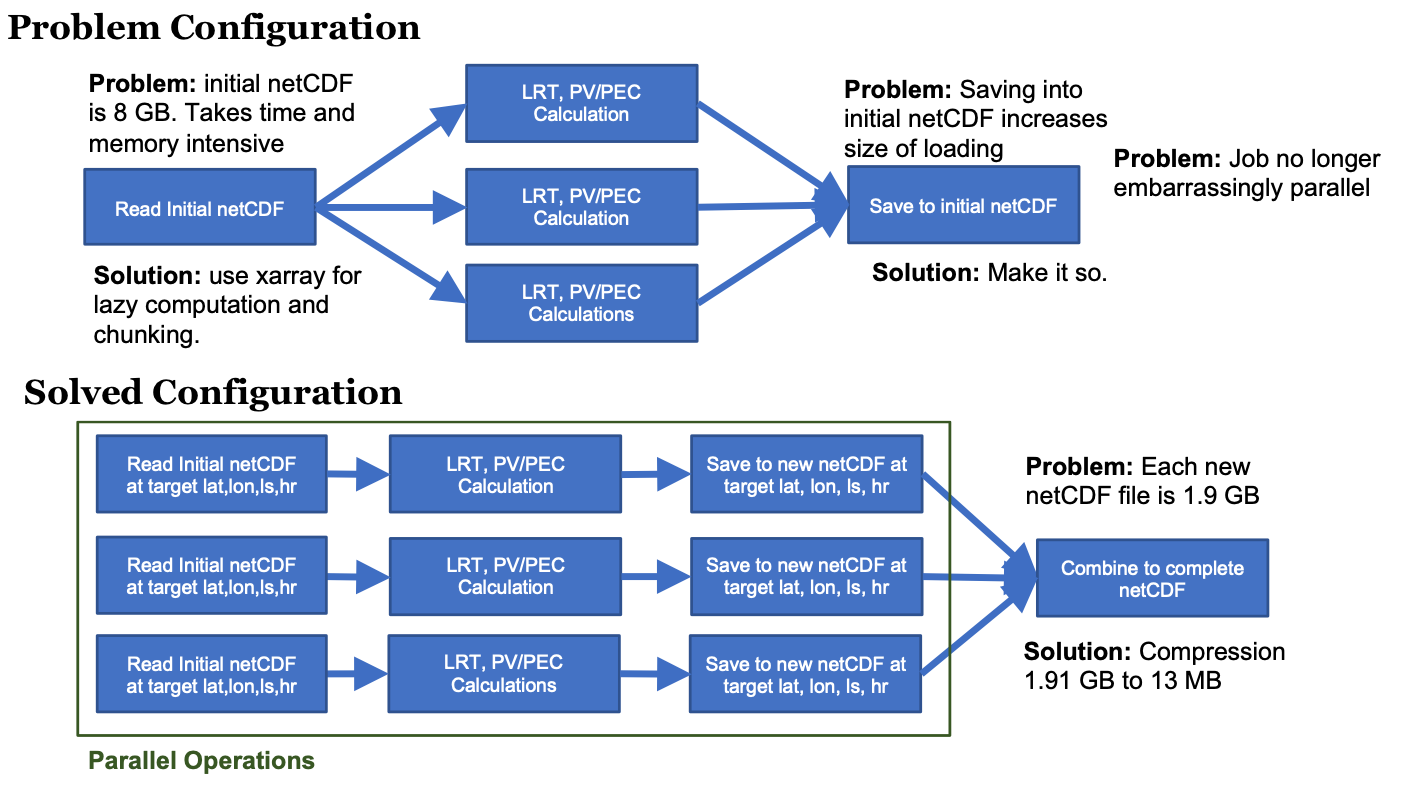}
    \caption{Initial (problem) and final (solution) configurations for the RedSun software on the UC Berkeley cluster.}
    \label{fig:PV1}
\end{figure}

The calculation of a single gridpoint's spectral flux (via \lstinline{libRadtran}) and the corresponding photovoltaic and photoelectrochemical production quantities ran for $\sim$5 minutes. Given the grid of 228475 geotemporalspatial points composed of 19 points of 10$^{\circ}$ latitude $\times$ 37 points of 10$^{\circ}$ longitude $\times$ 25 points of 15$^{\circ}$ areocentric longitude $\times$ 13 points of 2 (Martian) hours, a serial calculation would require 2.17 years. Wanting to avoid that lengthy calculation, we opted for an ``embarrassingly parallel'' computing method shown in Figure \ref{fig:PV1}. 
Since our computations require some initial or final communication (generally in the distribution and collection of data, then we call it nearly embarrassingly parallel. In parallel computing, an embarrassingly parallel workload or problem is one where little or no effort is needed to separate the problem into a number of parallel tasks. This is often the case where there is little or no dependency or need for communication between those parallel tasks, or for results between them. In the ideal case, all the sub-problems or tasks are defined before the computations begin and all the sub-solutions are stored in independent memory locations (variables, array elements). Thus, the computation of the sub-solutions is completely independent\footnote{\url{https://www.cs.iusb.edu/~danav/teach/b424/b424_23_embpar.html}}. 

Files were not constructed for grid-points that did not receive sunlight, and so the result was the storage of $\sim$150k \lstinline{.netCDF} files, each with a size of $\sim$4-5 MB. 

\subsection{Stitching}
The $\sim$150k singlepoint \lstinline{.netCDF} files were initially stitched across time dimensions of hours and areocentric longitude to produce $\sim$700 time series \lstinline{.netCDF} files, each for a different pair of latitudes and longitudes using the \lstinline{tcsh} scripts provided in Listing \ref{bill1} and \ref{bill2}. 

\begin{lstlisting}[language=bash, caption=Stitching Algorithm Part 1: Create Dynamic Links, label={bill1}]
#!/bin/tcsh -f
if ($#argv != 1) then
   echo "--> usage: csh " $0 " netcdf_file"
   exit
endif
set link = `ncdump -v ls,hr,lat,lon $argv[1] | sed -n '/^data:/,$p' | sort | paste -s -d" " - | awk '{printf("%s%03d%02d%s%02d%s%02d%s\n","ttlrecall_",$15,$3,"_",$11,"_",$7,".nc");}'`
ln -sv $argv[1] $link
\end{lstlisting}

\begin{lstlisting}[language=bash, caption=Stitching Algorithm Part 2: Assemble into Time Series, label={bill2}]
#!/bin/tcsh -f
set lat = minimum_lat_value
set lon = minimum_lon_value
while ($lat <= maximum_lat_value)
    set latv = `echo $lat | awk '{printf("%02d\n",$1)}'`
    while ($lon <= maximum_lon_value)
           set lonv = `echo $lon | awk '{printf("%02d\n",$1)}'`
           ncecat ttlrecall_*_{$lonv}_{$latv}.nc redsun_timeseries_{$lonv}_{$latv}.nc
           echo "Done: " $lonv $latv
@ lon++
     end
@ lat++
end
\end{lstlisting}

\subsection{Production Mapping}
The resultant timeseries \lstinline{.netCDF} files were then used for constructing the final maps of PV and PEC production. For each time series \lstinline{.netCDF} file, we began by calculating PV power $P$ and PEC production rate $\dot{m}$ via
\begin{align}
    P & = \Gamma\eta_{\text{pv}}\\
    \dot{m_{c}} &= \epsilon_{c}\Gamma\eta_{\text{pec}} = \frac{Z_{c}}{n_{c}V_{c}F}\Gamma\eta_{\text{pec}}
\end{align}
where $\Gamma$ is the solar flux in W/m$^{2}$ sourced from the \lstinline{MCD} data in \lstinline{StupidGrid.nc}, $\epsilon$ is the electrochemical equivalency factor, $\eta$ is the calculated PV/PEC efficiency, $Z$ is the molar mass, $n$ is the number of moles of electrons required to make one mole of the product, $F$ is the Faraday constant, and $V$ is the voltage. The $c$ term corresponds to the chemical of interest in the set of H$_{2}$, NH$_{3}$, and AA. The values used to produce the $\epsilon$ for each chemical is given in Table \ref{tab:prodpar}.

\begin{table}[h!]
\centering
\begin{tabular}{@{}llll@{}}
\toprule
Chemical & $n$ & $Z$      & $V$    \\ \midrule
H$_{2}$       & 2 & 2.016  & 1.23 \\
NH$_{3}$      & 6 & 17.031 & 1.17      \\
AA       & 8 & 60.052 & 1.09      \\ \bottomrule
\end{tabular}
\caption{Electrochemical equivalency factor parameters.}
\label{tab:prodpar}
\end{table}

We calculated the optimal sol-averaged 3-junction PV $P_{\text{opt}}$ and 2-junction PEC $\dot{m_{c}}_{\text{opt}}$ across all bandgap combinations given the form
\begin{align}
    P_{\text{opt}} = \max\left(\frac{1}{N}\int_{t_{2}}\int_{t_{1}}P_{ijk}dt_{1}dt_{2} : \forall i,j,k \in B_{1},B_{2},B_{3}\right)\\
    \dot{m}_{c,opt} = \max\left(\frac{1}{N}\int_{t_{2}}\int_{t_{1}}\dot{m}_{c,ij}dt_{1}dt_{2} : \forall i,j \in B_{1},B_{2}\right)
\end{align}
where $i,j,k$ are indices of bandgaps $B_{1},B_{2},B_{3}$, $t_{1}$ is the time variable across a sol ($\sim$24.616 hrs/sol), and $t_{2}$ is the time variable across a Martian year given as $N=688$ sols/year. 

Computationally, we began by converting our $L_{s}$ values to the sol number using an inverted Kepler problem with a function \lstinline{ls2sol} shown in Listing \ref{ls2sol}.

\begin{lstlisting}[language=Python, caption=Function for converting $L_{s}$ to sol number, label={ls2sol}]
def ls2sol(ls):
    N_s = 668.6
    ls_peri = 250.99
    t_peri = 485.35
    a = 1.52368
    e = 0.09340
    epsilon = 25.1919
    if (ls == 0).any():
        ls = .01
    nu = np.radians(ls) + 1.90258
    E = np.arctan((np.tan(nu/2))/(np.sqrt((1+e)/(1-e))))*2
    M = E - e*np.sin(E)
    Ds = (M/(2*np.pi))*N_s + t_peri
    if (Ds < 0).any():
        Ds = Ds + N_s
    if (Ds > N_s).any():
        Ds = Ds - N_s
    return(Ds)
\end{lstlisting}

The computational instance of calculations for 2J H$_{2}$ production is provided in Listing \ref{pointloop}.

\begin{lstlisting}[language=Python, caption=Function for calculating the optimal H$_{2}$ production rate, label={pointloop}]
def point_loop(file):
    sg = xr.open_dataset('StupidGridFull.nc', group='flux')
    ds = xr.open_dataset(file)
    lat = ds['lat'][0]
    lon = ds['lon'][0]
    G = np.zeros(len(ds['lon']))
    for ri in range(0,len(ds['lon'])):
        ls = ds['ls'][ri]
        hr = ds['hr'][ri]
        G[ri] = sg['flux_dw_sw'][lat,lon,ls,hr]
    lss = np.unique(ds['ls'])
    Z = 2.016
    n = 2
    F = 96485.33212
    V = 1.23
    sg = 0
    sols = np.zeros(len(lss))
    for i in range(0,len(lss)):
        sols[i] = ls2sol(lss[i]*15)
    hrs = np.arange(0,25,2)
    vals = np.zeros(13)
    try: 
        P = G[:, np.newaxis, np.newaxis] * ds['j2_etaPEC_H2_2bg'] * 0.01 * Z/(n*F*V)
        zz = np.zeros((len(lss),len(ds['j2-bg1']),len(ds['j2-bg2'])))
        for i in range(0,len(lss)):
            hr_int = np.where(ds['ls']==lss[i])
            inds = np.array(ds['hr'][hr_int])
            for j in range(0,len(ds['j2-bg1'])):
                for k in range(0,len(ds['j2-bg2'])):
                    y = P[:,j,k][hr_int]
                    for m in range(0,len(inds)):
                        vals[inds[m]] = y[m]
                    z = np.trapz(vals*60*60,x=hrs*1.02569)
                    zz[i,j,k] = z
        z = np.zeros((len(ds['j2-bg1']),len(ds['j2-bg2'])))
        for j in range(0,len(ds['j2-bg1'])):
            for k in range(0,len(ds['j2-bg2'])):
                y = zz[:,j,k]
                z[j,k] = np.trapz(y,x=sols)
        j2h2 = np.max(z)
        j2h2i = np.unravel_index(np.argmax(z),np.shape(z), order='C')
        h2 = j2h2 * (1/688)
        bg1 = ds['j2-bg1'][j2h2i[0]]
        bg2 = ds['j2-bg2'][j2h2i[1]]
        return([[lat,lon,0],[h2,bg1,bg2]])
\end{lstlisting}

The results from the calculation of the optimal sol-averaged 3-junction PV $P_{\text{opt}}$ and 2-junction PEC $\dot{m_{c}}_{\text{opt}}$ and their corresponding bandgap combination were again saved as \lstinline{.netCDF} files with dimensions of latitude and longitude.

\begin{figure}[h!]
\centering
\includegraphics[width=0.65\textwidth]{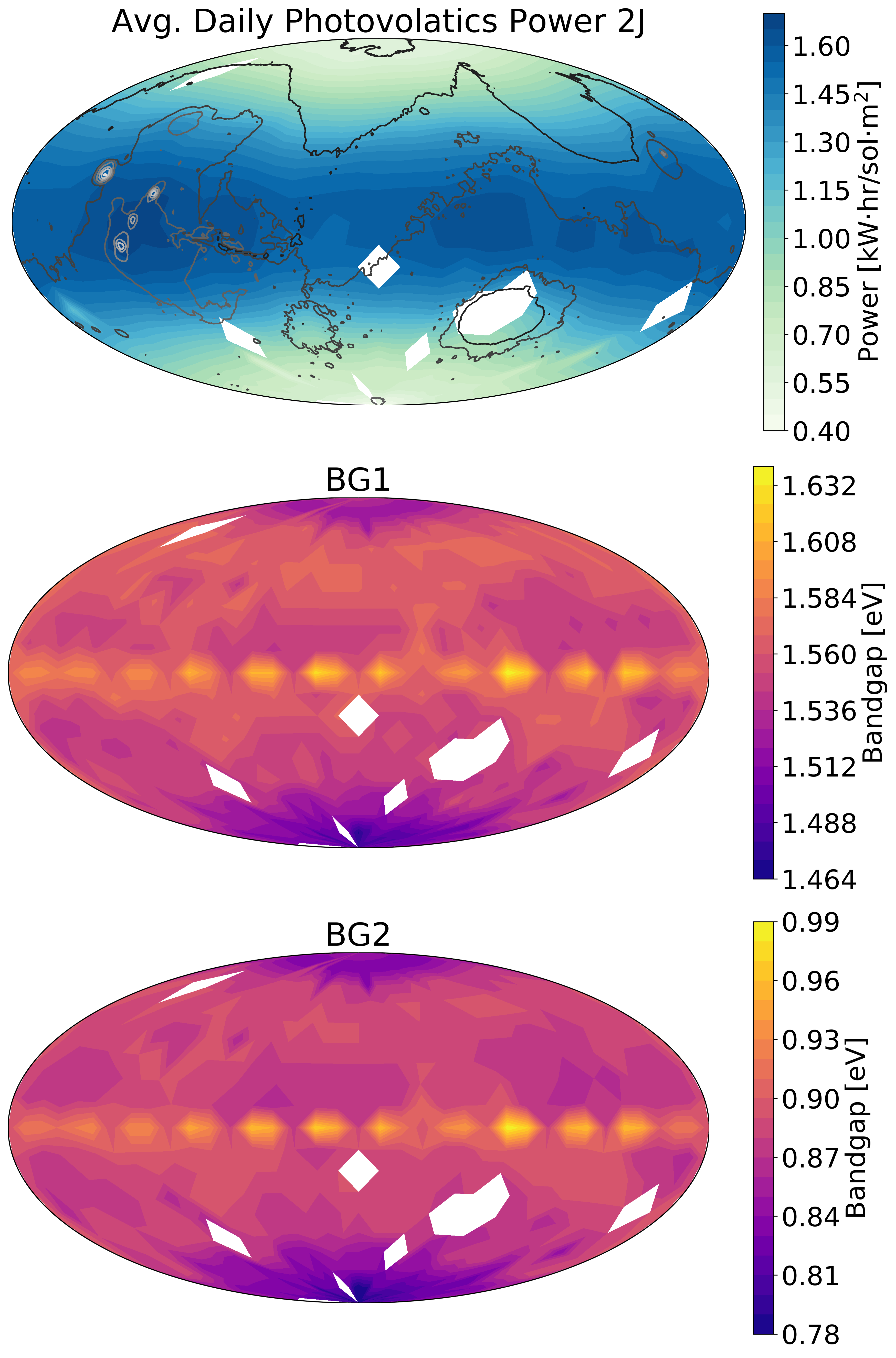}
\caption{Two Junction Photovoltaic Power Production and Optimal Bandgaps distributed over the Martian Grid}
\label{2jpv}
\end{figure}

\begin{figure}[h!]
\centering
\includegraphics[width=0.65\textwidth]{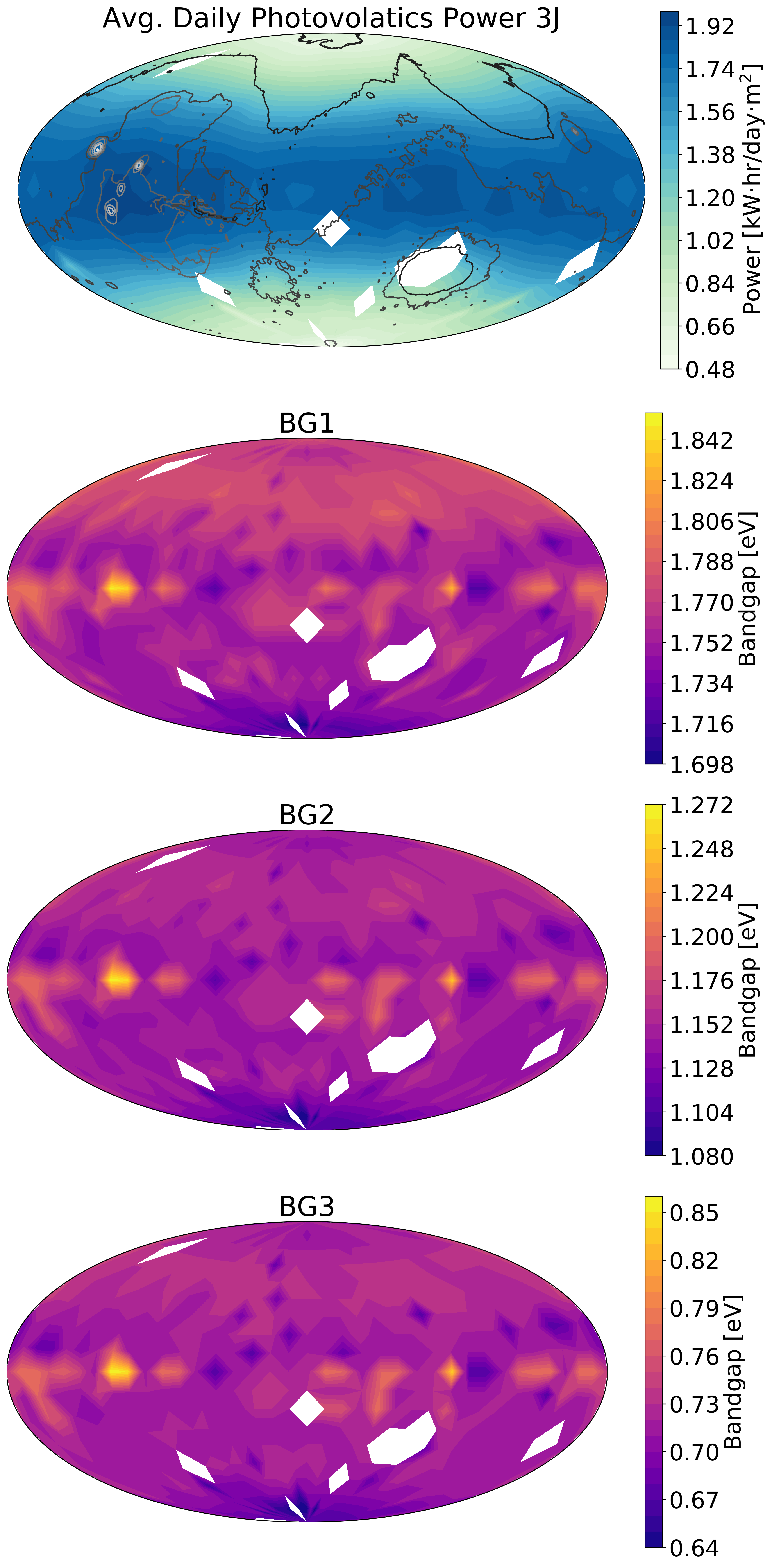}
\caption{Three Junction Photovoltaic Power Production and Optimal Bandgaps distributed over the Martian Grid}
\label{3jpv}
\end{figure}

\begin{figure}[h!]
\centering
\includegraphics[width=0.65\textwidth]{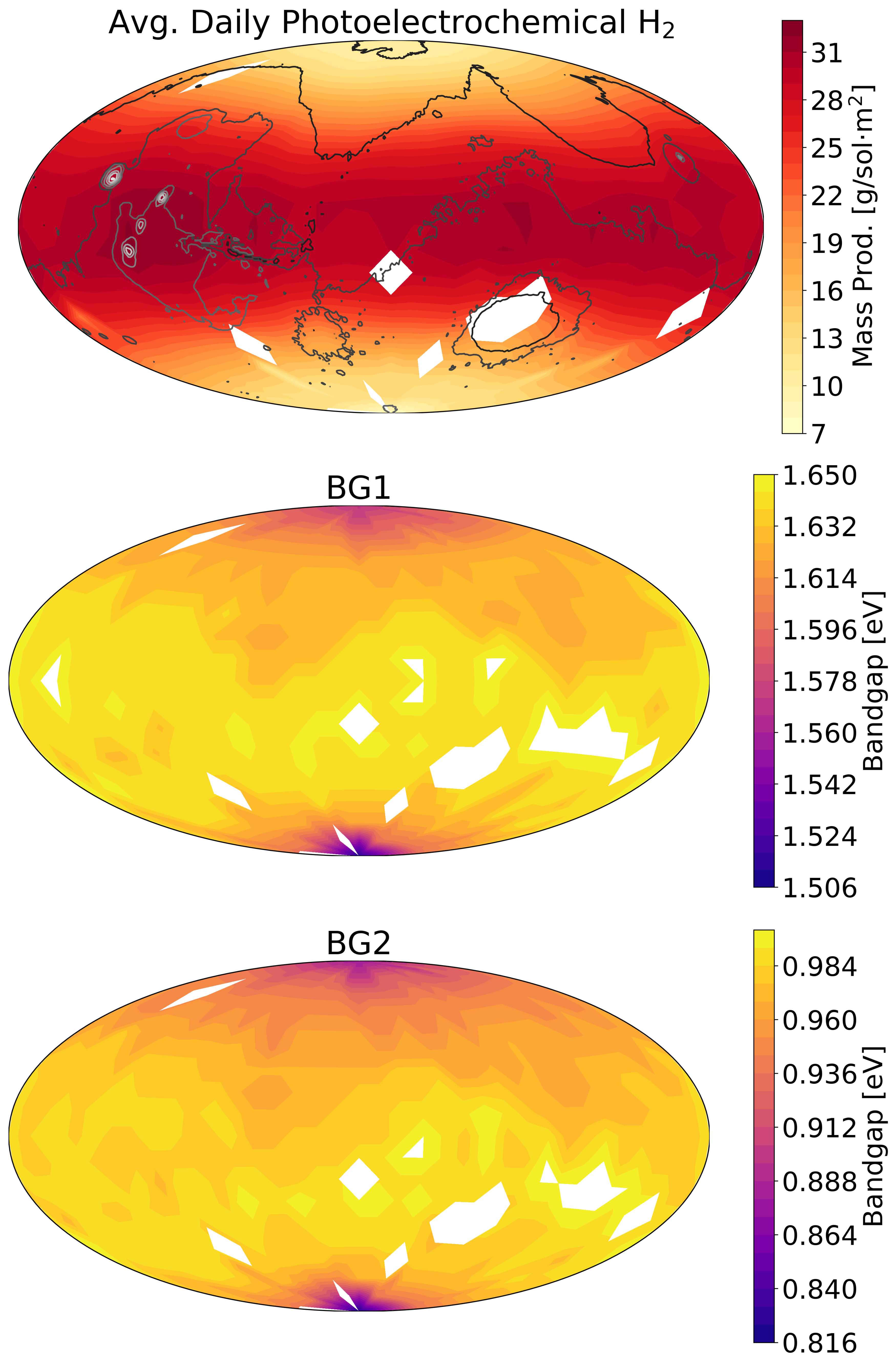}
\caption{Two Junction Photoelectrochemical H$_{2}$ Production and Optimal Bandgaps distributed over the Martian Grid}
\label{2jh2}
\end{figure}

\begin{figure}[h!]
\centering
\includegraphics[width=0.95\textwidth]{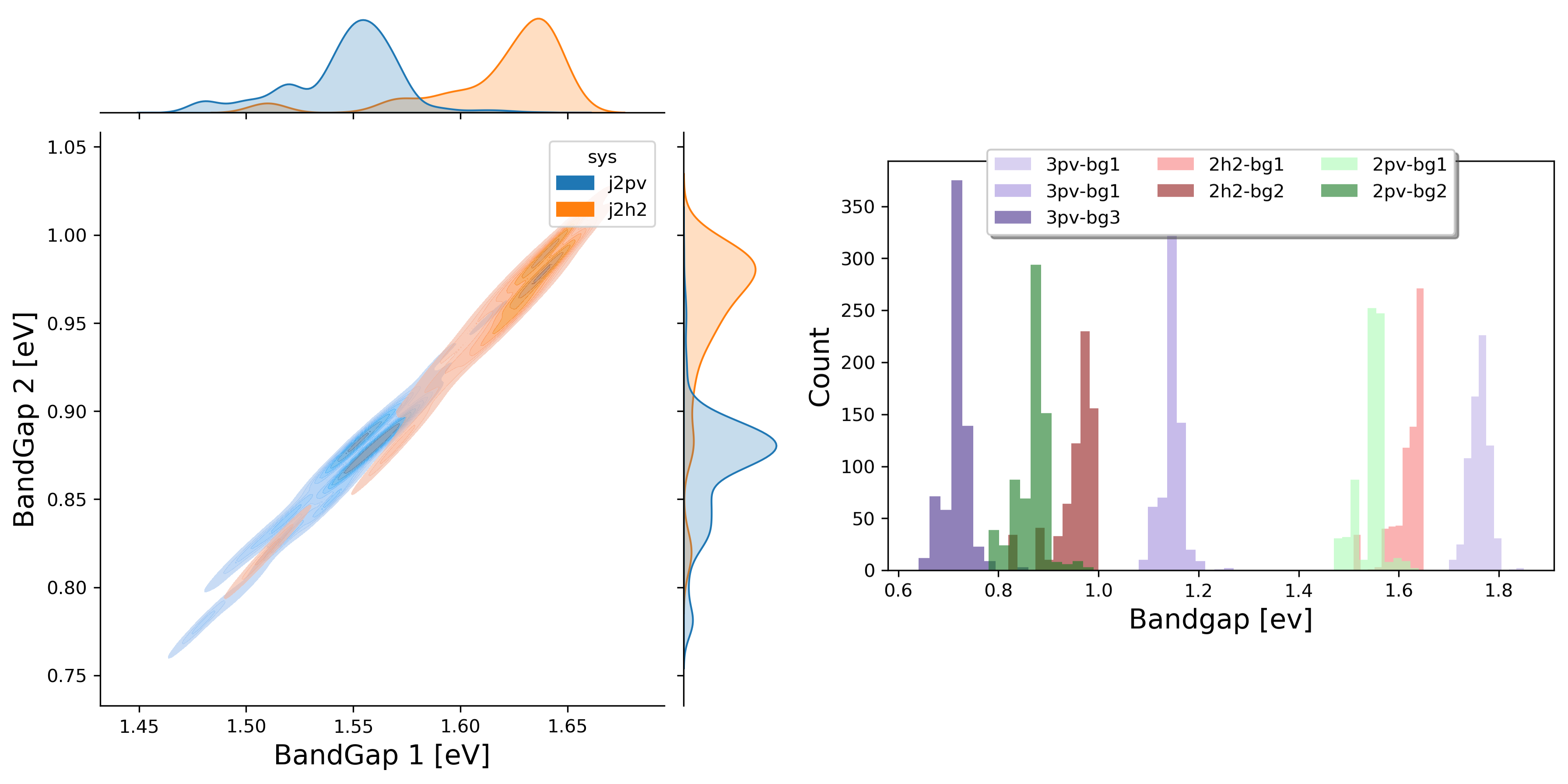}
\caption{Optimal Bandgap Distributions.}
\label{bgdist}
\end{figure}

The resulting PV power and PEC production for H$_{2}$ is provided in Figure \ref{2jpv}-\ref{2jh2} with the corresponding Bandgaps distributions over the Martian grid. The distribution of bandgaps are provided in Figure \ref{bgdist}.

\begin{table}[h!]
\centering
\begin{tabular}{@{}|l|l|l|@{}}
\toprule
\textbf{Commodity} &
  \textbf{\begin{tabular}[c]{@{}l@{}}Best efficiency at\\ averaged solar noon\end{tabular}} &
  \textbf{\begin{tabular}[c]{@{}l@{}}Best production \\ over a year\end{tabular}} \\ \midrule
\multirow{3}{*}{Power (PV,  3-junction)} & Top: 1.77 eV    & Top: 1.83 eV     \\ \cmidrule(l){2-3} 
                                         & Middle: 1.16 eV & Middle: 1.16 eV  \\ \cmidrule(l){2-3} 
                                         & Bottom: 0.72 eV & Bottom: 0. 67 eV \\ \midrule
\multirow{2}{*}{H$_{2}$  (PEC, 2-junction)}   & Top: 1.64 eV    & Top: 1.77 eV     \\ \cmidrule(l){2-3} 
                                         & Bottom: 0.95 eV & Bottom: 0.83 eV  \\ \bottomrule
\end{tabular}
\caption{Comparison of optimal bandgaps for different optimization strategies}
\label{tab:bg}
\end{table}

\subsection{Missing Location Values}
We were able to complete the calculations for $\sim$97\% of the 228475 geospatial points across the Martian grid. We found that $\sim$6000 of these points could not be completed due to a number of issues our method of using libRadtran for Mars-based calculations. Upon inspection, we found that the missing values were generally concentrated in areas with very low elevation below the Martian datum. Further inspection confirmed that the issues in resolving the radiative transfer were caused by errors in interpolation by the solver for the gas concentrations below the datum. However, these $\sim$2\% of missing values do not prevent us from offering a meaningful analysis.

\cleardoublepage
\section{Technoeconomic Calculations}
\subsection{Primary Power and Energy Demands}
\begin{figure}[b!]
\centering
\includegraphics[width=0.8\textwidth]{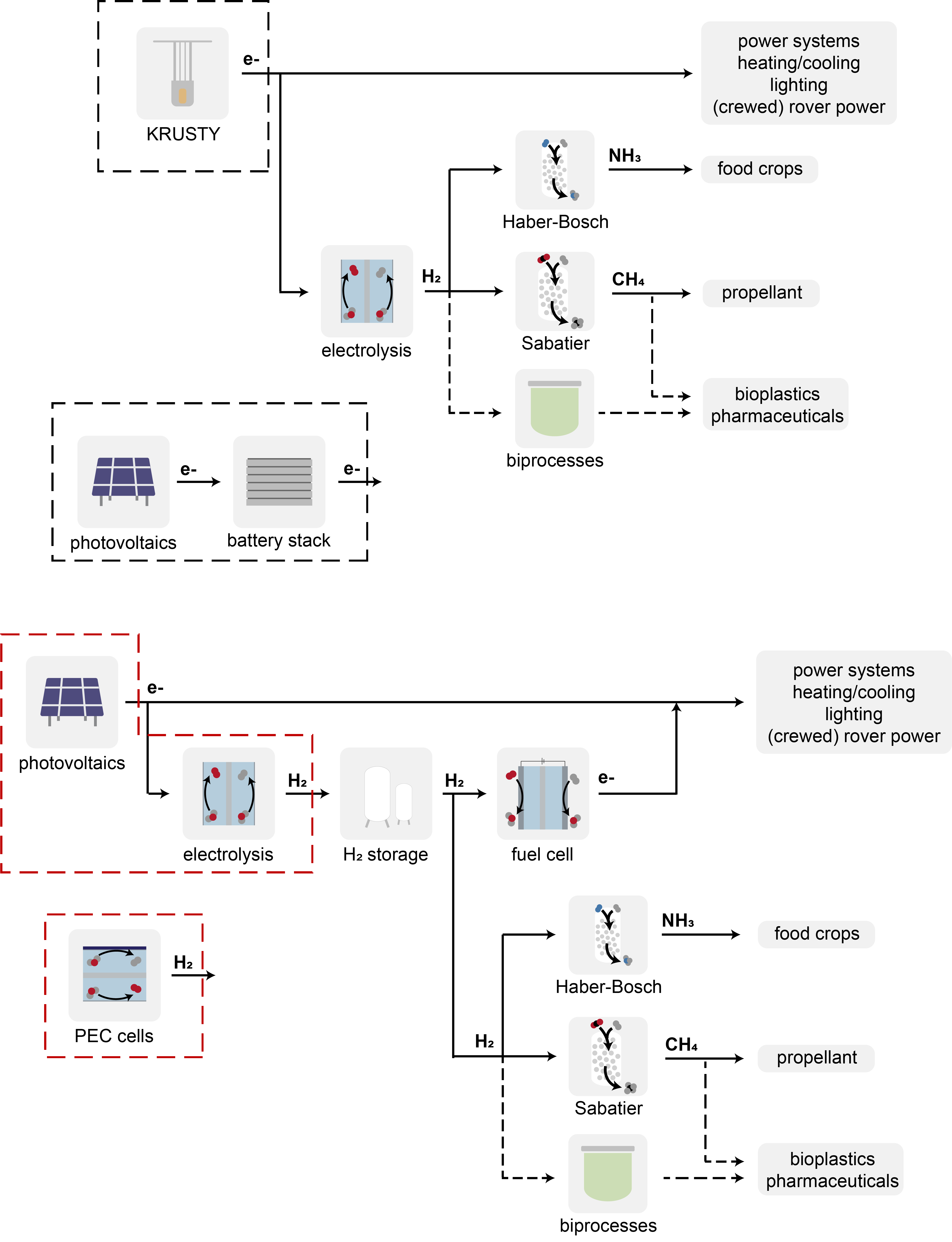}
\caption{\textbf{Power generation systems options.} Habitat power systems and ammonia, propellant, and bioplastics production can be powered by nuclear power generation (KRUSTY), photovoltaics with battery storage (PV+B), photovoltaics with H2 energy storage from hydrolysis (PV+E), or photoelectrochemical H2 generation and storage (PEC).}
\label{processdiagram}
\end{figure}

We consider four different power production and energy storage scenarios for comparison (Fig. \ref{processdiagram}): (1) Nuclear power generation with the Kilopower system; (2) Photovoltaic power generation with battery energy storage; (3) Photovoltaic power generation with compressed H$_{2}$ energy storage, and (4) Photoelectrochemical H$_{2}$ generation with compressed H$_{2}$ energy storage.

In all cases, power and/or energy demand is driven by continuous power required for habitat operations, including lighting, heating/cooling, pressurization, power draw for ISRU processes, and power draw for rover travel, and by materials demand for ISRU manufacturing. We assume that ammonia, methane, and plastics are produced using H$_{2}$ as the starting material (along with N$_{2}$ and CO$_{2}$ sourced from the atmosphere), which we use to calculate power demands based on water electrolysis to produce H$_{2}$. We note that methane could be diverted for bioprocess production (dashed lines in Fig. \ref{processdiagram}), although we don’t explicitly consider this scenario here since it would not change the relative mass requirements of the four systems we consider.

\begin{figure}[t!]
\centering
\includegraphics[width=0.65\textwidth]{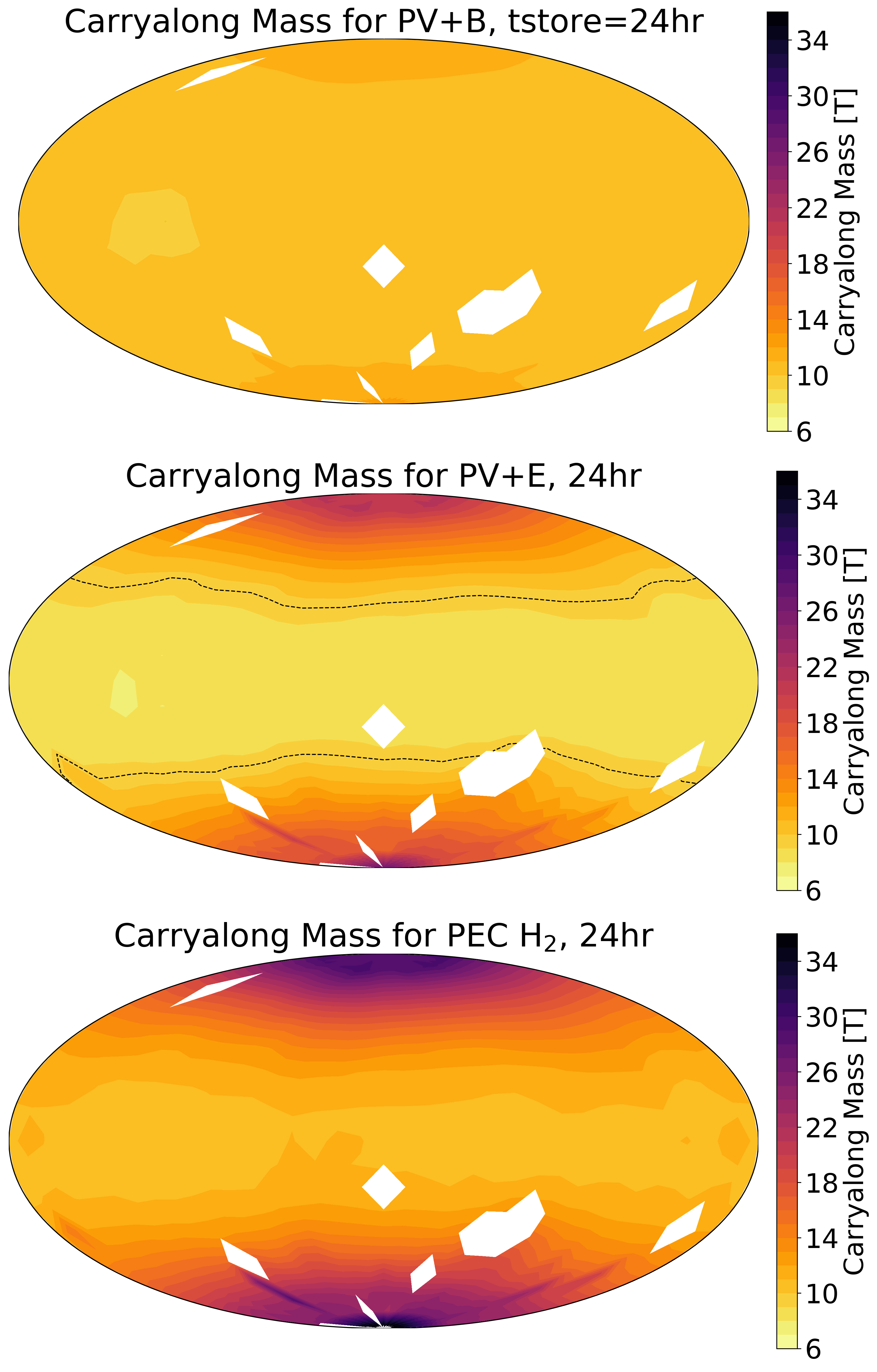}
\caption{\textbf{Carry-along mass for different power generation scenarios.} Carry-along mass across the Martian surface for PV+B, PV+E, and PEC power generation systems. PV+B and PEC systems cannot reach parity with nuclear power generation in terms of carry along mass (no locations at which the projected mass of the PV+B or PEC systems is less than the projected mass of the nuclear system).}
\label{carryalong}
\end{figure}

To compare the carry-along mass necessary for each system, we include the mass of elements unique to or uniquely sized for a given energy supply scenario. For example, we consider the mass of photovoltaic cells because the area of cells necessary to power the habitat and ISRU manufacturing will be different depending on the strategy for energy storage. However, we don’t include the mass of the Sabatier reactor for methane production, since this mass will be equivalent regardless of the upstream processes producing H$_{2}$ and collecting CO$_{2}$ from the atmosphere. In this way, we can determine the mass contributions only of the uniquely necessary components for each energy supply scenario. The carry along masses are provided in Figure \ref{carryalong}.

\subsubsection{Nuclear Power}
\label{notenuc}
Power derived from the Kilopower nuclear reactor system is fed directly to habitat power systems and to an electrolyzer producing H$_{2}$ for ISRU manufacturing. Hence, the power draw is given by:
\begin{align}
    P_{\text{K}} &= P_{\text{Hab}} + \alpha_{\text{E}}\left(\dot{N}\alpha_{\text{HB}} + \dot{M}\alpha_{\text{S}} + \dot{B}\alpha_{\text{HB}}\right)\\
    P_{\text{K}} &= P_{\text{Hab}} + \alpha_{\text{E}}\Lambda
\end{align}
where $P_{\text{K}}$ is the total power draw for Kilopower nuclear reactor system, $P_{\text{Hab}}$ is the power draw for the habitat, $\alpha_{\text{E}}$ is the energy demand per unit of H$_{2}$ produced for the electrolyzer, $\dot{N}$ is the ammonia demand rate, $\dot{M}$ is the methane demand rate, $\dot{B}$ is the bioplastic demand rate, and $\alpha_{i}$ is the conversion factor between, e.g., the ammonia demand rate and the H$_{2}$ demand rate for the Haber-Bosch process. We also define $\Lambda=\dot{N}\alpha_{\text{HB}} + \dot{M}\alpha_{\text{S}} + \dot{B}\alpha_{\text{HB}}$.

The carry-along mass requirements for this scenario is given by
\begin{align}
    M_{\text{K}} &= \frac{P_{\text{K}}}{p_{\text{K}}} + \frac{\Lambda}{p_{\text{E}}}
\end{align}
where $p_{\text{K}}$ is the specific power of the Kilopower reactor (6.25 W/kg) and $p_{\text{E}}$ is the specific productivity of the electrolyzer (kg H$_{2}$/h/kg).

\subsubsection{Photovoltaic power with battery energy storage (PV+B)}
\label{notebat}
Power generated by photovoltaic cells can be transferred either directly to power-drawing systems (habitat systems, water electrolysis) or diverted to battery stacks for storage to enable continuous operation either at night or during low-sunlight days (due to high dust conditions). We define the fraction of power supplied directly to power systems as $\chi$, which, for photovoltaic systems, can be thought of as the fraction of the day that solar cells produce equal or more power than what is consumed by power-drawing systems. Unless otherwise stated, we assume in our calculations $\chi=1/3$. Hence, the total power draw for the PV+B system is given by:
\begin{align}
    P_{\text{PV+B}} &= \chi P_{\text{Hab}} + \frac{1-\chi}{\eta_{\text{B}}}P_{\text{Hab}} + \chi\alpha_{\text{E}}\Lambda + \frac{1-\chi}{\eta_{\text{B}}}\alpha_{\text{E}}\Lambda
\end{align}
where $P_{\text{PV+B}}$ is the total power draw for the PV+B system and $\eta_{\text{B}}$ is the energy efficiency of the battery storage system. More compactly,
\begin{align}
    P_{\text{PV+B}} &= \left(\chi + \frac{1-\chi}{\eta_{\text{B}}}\right)\left(P_{\text{Hab}}+\alpha_{\text{E}}\Lambda\right)
\end{align}
The carry-along mass required for the PV+B scenario is given by
\begin{align}
    M_{\text{PV+B}} &= \frac{P_{\text{PV+B}}}{p_{\text{PV}}}+\frac{\left(P_{\text{Hab}}+\alpha_{\text{E}}\Lambda\right)}{e_{\text{B}}}t_{\text{store}} + \frac{\Lambda}{p_{\text{E}}}
\end{align}
where $p_{\text{PV}}$ is the specific power of photovoltaic cells, $t_{\text{store}}$ is the desired back-up power availability time, and $e_{\text{B}}$ is the specific energy of the battery stack (units of energy per mass).

\begin{table}[h!]
\centering
\begin{tabular}{@{}llll@{}}
\toprule
\textbf{Parameter} & \textbf{Value} & \textbf{Unit} & \textbf{Reference} \\ \midrule
\multicolumn{4}{l}{\textbf{Power and Material Demands}}\\ \midrule
 $P_{\text{Hab}}$ & 40 & kW & Note \ref{note1.1} \\
 $\dot{N}$& 8.33$\times$10$^{-3}$ & kg h$^{-1}$ & Note \ref{note1.2} \\
 $\dot{M}$& 0.61 & kg h$^{-1}$ & Note \ref{note1.3} \\
 $\dot{B}$& 0.1 & kg h$^{-1}$ & Note \ref{note1.4}\\ \midrule
\multicolumn{4}{l}{\textbf{Conversion Factors}}\\ \midrule
 $\alpha_{\text{HB}}$ & 0.196 & kgH$_{2}$ kgNH$_{3}^{-1}$ & Note \ref{note1.2} \\
 $\alpha_{\text{S}}$ & 0.554 & kgH$_{2}$ kgCH$_{4}^{-1}$ & Note \ref{note1.3} \\
 $\alpha_{\text{BP}}$ & 0.155 & kgH$_{2}$ kgAA$^{-1}$ & Note \ref{note1.4} \\
 $\alpha_{\text{E}}$ & 54.13 & kWh kgH$_{2}^{-1}$ & Note \ref{note1.5} \\
 $\alpha_{\text{FC}}$ & 0.064 & kgH$_{2}$ kWh$^{-1}$ & Note \ref{note1.5} \\
 $\alpha_{\text{HS}}$ & 3.39 & kWh kgH$_{2}^{-1}$ & Note \ref{note1.5}\\ \midrule
\multicolumn{4}{l}{Power\cite{Hannan2018State-of-the-artRecommendations} and Energy Density\cite{Eftekhari2017EnergyResearch}}\\ \midrule
 $p_{\text{K}}$ & 6.25$\times$10$^{-3}$ & kW kg$^{-1}$ & Note \ref{notenuc} \\
 $\eta_{\text{B}}$ & 80 & \% & Note \ref{notebat} \\
 $p_{\text{E}}$ & 1.14$\times$10$^{-2}$ & kgH$_{2}$ h$^{-1}$ kg$^{-1}$ & Note \ref{note1.5} \\
 $e_{\text{B}}$ & 0.16 & kWh kg$^{-1}$ & Note \ref{notebat} \\
 $p_{\text{FC}}$ & 0.365 & kW kg$^{-1}$ & Note \ref{note1.5} \\
 $e_{\text{HS}}$ & 7.18$\times$10$^{-2}$ & kgH$_{2}$ kg$^{-1}$ & Note \ref{note1.5}\\ \midrule
\multicolumn{4}{l}{\textbf{Solar Cell Array Mess}}\\ \midrule
 $M_{\text{PV}}$ & 2 & kg m$^{-2}$ & Note \ref{note1.6} \\
 $M_{\text{PEC}}$ & 2.4 & kg m$^{-2}$ & Note \ref{note1.6}\\ \midrule
\multicolumn{4}{c}{\textbf{Other Parameters}}\\ \midrule
 $\chi$& 0.33 & -- & Assumed \\
 $t_{\text{store}}$ & 24.6 & h & Assumed \\ \bottomrule
\end{tabular}
\caption{}
\label{tab:demand}
\end{table}

\subsubsection{Photovoltaic power with H$_{2}$ energy storage}
In this scenario, power generated by photovoltaic cells can either be directly fed to habitat systems or to an electrolyzer, which produces H$_{2}$ for consumption in ISRU manufacturing and for consumption by fuel cells the supply power to the habitat and other demands when direct power cannot (e.g., at night). Here, the total power demand for the system is given by
\begin{align}
    P_{\text{PV+E}} = \chi P_{\text{Hab}} + \alpha_{\text{E}}\dot{m}_{\text{H}_{2}}
\end{align}
where $P_{\text{PV+E}}$ is the total power draw for the PV+E system and $\dot{m}_{\text{H}_{2}}$ is the flow rate of H$_{2}$ necessary to support the remaining system requirements. This flow rate is written as
\begin{align}
    \dot{m}_{\text{H}_{2}} &= \frac{(1-\chi)P_{\text{Hab}}\alpha_{\text{FC}}+\Lambda}{1-\alpha_{\text{HS}}\alpha_{\text{FC}}}
\end{align}
where $\alpha_{\text{FC}}$ is the H$_{2}$ consumed per unit of energy produced by the fuel cell and $\alpha_{\text{HS}}$ is the energy consumed per unit of H$_{2}$ stored by the H$_{2}$ storage tanks (driven by compression of H$_{2}$).

The carry-along mass required for the PV+E scenario is given by
\begin{align}
    M_{\text{PV+E}} &= \frac{P_{\text{PV+E}}}{p_{\text{PV}}}+ \frac{\dot{m}_{\text{H}_{2}}}{p_{\text{E}}} + \frac{P_{\text{Hab}}+\alpha_{\text{HS}}\dot{m}_{\text{H}_{2}}}{p_{\text{FC}}} + \frac{(P_{\text{Hab}}\alpha_{\text{FC}}+\Lambda)t_{\text{store}}}{e_{\text{HS}}}
\end{align}
where $p_{\text{FC}}$ is the specific power of the fuel cell and $e_{\text{HS}}$ is the specific mass of the H$_{2}$ storage tanks (in units kgH$_{2}$/kg$_{\text{tank}}$).

\subsubsection{Photoelectrochemical (PEC) H$_2$ generation with H$_{2}$ energy storage}
This scenario uses an H$_{2}$ demand as opposed to a power demand to size the PEC array. The total H$_{2}$ demand rate is given by
\begin{align}
    \dot{m}_{\text{H}_{2}} &= \frac{P_{\text{Hab}}\alpha_{\text{FC}}+\Lambda}{1-\alpha_{\text{HS}}\alpha_\text{FC}}
\end{align}

The carry-along mass required for the PEC scenario is given by
\begin{align}
    M_{\text{PEC}} &= \frac{\dot{m}_{\text{H}_{2}}}{m_{\text{PEC}}} + \frac{P_{\text{Hab}}+\alpha_{\text{HS}}\dot{m}_{\text{H}_{2}}}{p_{\text{FC}}} + \frac{(P_{\text{Hab}}\alpha_{\text{FC}}+\Lambda)t_{\text{store}}}{e_{\text{HS}}}
\end{align}
where $m_{\text{PEC}}$ is the specific productivity (kgH$_{2}$/h/kg) of PEC cells. All parameters for these calculations are compiled in Table \ref{tab:demand}.

\subsection{Secondary Power and Energy Demands}
\subsubsection{Habitat Power Demand}
\label{note1.1}
Continuous power demand estimates for a Martian habitat range between 4 and $\sim$100 kW. We use 40 kW as a baseline value following the NASA Baseline Values and Assumptions Document (BVAD)\cite{Anderson2018}. This value includes ISRU power demands, including for crop growth, so we only calculated additional power demands for H$_{2}$ production for the ISRU processes considered.

\subsubsection{Ammonia Demand}
\label{note1.2}
To calculate an upper-bound ammonia demand, we followed the optimization strategy by Do \textit{et al}. assuming no recycling of nitrogen via urea recovery\cite{Do2016AnAnalysis}. Briefly, we assumed that the metabolic demands for six crew members would be met entirely by food crops grown in hydroponic chambers. We used values from the BVAD and related literature to calculate nitrogen demand per nutrient availability for a given crop\cite{Anderson2018,Wheeler2003CropProject}. The optimization function was defined to balance minimization of area necessary for crop growth with maximization of crop variability for human morale as 
\begin{align}
    f &= w_{1}\sum_{i}A_{i} + w_{2}\sigma(\boldsymbol{A})\\
    &s.t.: \sum_{i}A_{i}r_{i}x_{i,j} > X_{j}
\end{align}
where $f$ is the optimization function, $A_i$ is the growth area for crop $i$, $\sigma$ is the standard deviation of the vector of crop areas ($\boldsymbol{A}$), $r_i$ is the static growth rate, $x_{i,j}$ is the nutritional content of crop i for nutrient $j$, and $X_j$ is the crew member demand for nutrient $j$. The relative weights $w_1$ and $w_2$ are related by \begin{align}
    w_2&=1-w_1
\end{align}
and $w_1$ was varied between 0 and 1. Using $w_1=0.25$, all 5 crops we considered (soybeans, wheat, lettuce, potatoes, peanuts) were included, resulting in a total crop growth area of $\sim$421 m$^{2}$ and an ammonia demand of $\sim$205 g/sol, which we converted to 8.33 g/h for consistent units in Table \ref{tab:demand}. The nitrogen demand ranged between $\sim$285 g/sol and $\sim$194 g/sol for $0<w_1<1$. 

We assume ammonia is produced via the Haber-Bosch process with the characteristic reaction
\begin{align}
    \text{N}_2+3\text{H}_2\rightarrow 2\text{NH}_3
\end{align}
Hence, the H$_{2}$:NH$_{3}$ conversion factor is 0.196 kgH$_{2}$/kgNH$_{3}$ assuming 90\% conversion of H$_{2}$.

\subsubsection{Methane Demand}
\label{note1.3}
Resupply and crew member return to Earth from Mars will require that interplanetary transit vehicles can be refueled on Mars. We use the estimate by Kleinhenz and Paz\cite{Kleinhenz2017AnVehicle} that such refueling requires 6978 kgCH$_{4}$ produced every 480 sols, corresponding to a CH$_{4}$ production rate of 0.61 kg/h. We assume this methane is produced via the Sabatier reaction:
\begin{align}
    \text{CO}_{2}+4\text{H}_2\rightarrow\text{CH}_4+2\text{H}_2\text{O}
\end{align}
resulting in an H$_{2}$:CH$_{4}$ conversion factor of 0.554 kg H$_{2}$/kgCH$_{4}$ assuming 90\% conversion efficiency.

\subsubsection{Bioplastics and Biopharmaceutical Demand}
\label{note1.4}
Bioplastics and pharmaceutical demands for a Martian habitat are not well-defined in the literature. For a system where 50\% of spare parts necessary for a habitat are generated via additive manufacturing based on ISRU, Owens \textit{et al.} estimated that 9800 kg of spare parts mass would be necessary over 260 months (an extremely long duration with multiple resupplies and crew member exchanges)\cite{Owens2015BenefitsMars} Assuming these spares are generated from bioplastics, which are in turn produced from acetic acid at 50\% yield by C$_{2}$ feedstock-utilizing microorganisms\cite{10.3389/fspas.2021.711550}, this corresponds to $\sim$0.1 kg/h acetic acid demand. We assume acetic acid is produced by acetogens with a molar ratio of 4.2:1 (corresponding to 95\% of H$_{2}$ reducing power diversion to acetic acid production, a common value for acetogens), this corresponds to an H$_{2}$:CH$_{3}$COOH ratio of 0.155 kgH$_{2}$/kg CH$_{3}$COOH assuming 90\% conversion.

Pharmaceutical demand is not expected to exceed 1 g/sol, so we neglect this amount for the purposes of our calculations here.

\subsubsection{Water electrolyzer, H$_{2}$ fuel cell, and H$_{2}$ storage systems}
\label{note1.5}
Water electrolysis and H$_{2}$ fuel cell power demands are based on commercially available, low-weight fuel cell systems designed for transportation vehicles\footnote{G-HFCS-6kW Hydrogen Fuel Cell Power Generator (Fuel Cell Store, Product Code: 1035012)}. The electrolyzer requires 54.13 kWh/kgH$_{2}$, while the fuel cell requires 0.064 kgH$_{2}$/kWh. We assume H$_{2}$ storage is accomplished with Type IV compression chambers at 350 bar, which stores H$_{2}$ at 20.77 kgH$_{2}$/m$^{3}$ with a tank mass of 289.23 kg/m$^{3}$, corresponding to a H$_{2}$ storage density of 0.0718 kgH$_{2}$/kg\cite{DiProfio2009ComparisonEvaluations,Barthelemy2017HydrogenPerspectives}. For these systems, 3.39 kWh/kgH$_{2}$ is required to compress H$_{2}$ to 350 bar, which we account for in the total power demand\cite{DiProfio2009ComparisonEvaluations}.

\subsubsection{Solar Cell Array Mass}
\label{note1.6}
Commercial low-weight, flexible solar cell arrays have an installed mass of 2.0 kg/m$^{2}$\footnote{MiaSolé Flex-03W Series Module with adhesive}. We are not aware of similarly commercial PEC arrays, so we assume that the installed mass is driven primarily by the absorber material as opposed to the catalyst layers or ion exchange membrane. We therefore estimate an installed mass of 2.4 kg/m$^{2}$ by assuming the absorber and housing components comprise 80\% of the installed mass.

\cleardoublepage
\bibliography{references}

\begin{thebibliography}{10}
\urlstyle{rm}
\expandafter\ifx\csname url\endcsname\relax
  \def\url#1{\texttt{#1}}\fi
\expandafter\ifx\csname urlprefix\endcsname\relax\def\urlprefix{URL }\fi
\expandafter\ifx\csname doiprefix\endcsname\relax\def\doiprefix{DOI: }\fi
\providecommand{\bibinfo}[2]{#2}
\providecommand{\eprint}[2][]{\url{#2}}

\bibitem{Bingham2003}
\bibinfo{author}{Bingham, S.~J.} \emph{et~al.}
\newblock \bibinfo{title}{{The Mars Climate Database}}.
\newblock \bibinfo{type}{Tech. Rep.} (\bibinfo{year}{2003}).

\bibitem{Patel2004AnnualSurface}
\bibinfo{author}{Patel, M.~R.} \emph{et~al.}
\newblock \bibinfo{journal}{\bibinfo{title}{{Annual solar UV exposure and
  biological effective dose rates on the Martian surface}}}.
\newblock {\emph{\JournalTitle{Advances in Space Research}}}
  \textbf{\bibinfo{volume}{33}}, \bibinfo{pages}{1247--1252}
  (\bibinfo{year}{2004}).

\bibitem{Vicente-Retortillo2015ASurface}
\bibinfo{author}{Vicente-Retortillo, A.}, \bibinfo{author}{Valero, F.},
  \bibinfo{author}{Vazquez, L.} \& \bibinfo{author}{Martinez, G.~M.}
\newblock \bibinfo{journal}{\bibinfo{title}{{A model to calculate solar
  radiation fluxes on the Martian surface}}}.
\newblock {\emph{\JournalTitle{Journal of Space Weather and Space Climate}}}
  \textbf{\bibinfo{volume}{5}}, \bibinfo{pages}{A33} (\bibinfo{year}{2015}).

\bibitem{Mayer2005}
\bibinfo{author}{Mayer, B.} \& \bibinfo{author}{Kylling, A.}
\newblock \bibinfo{journal}{\bibinfo{title}{{The libRadtran software package
  for radiative transfer calculations-description and examples of use}}}.
\newblock {\emph{\JournalTitle{Atmospheric Chemistry and Physics}}}
  \textbf{\bibinfo{volume}{5}}, \bibinfo{pages}{1855--1877}
  (\bibinfo{year}{2005}).

\bibitem{Emde2016The1}
\bibinfo{author}{Emde, C.} \emph{et~al.}
\newblock \bibinfo{journal}{\bibinfo{title}{{The libRadtran software package
  for radiative transfer calculations (version 2.0. 1)}}}.
\newblock {\emph{\JournalTitle{Geoscientific Model Development}}}
  \textbf{\bibinfo{volume}{9}}, \bibinfo{pages}{1647--1672}
  (\bibinfo{year}{2016}).

\bibitem{Wiscombe1980ImprovedAlgorithms}
\bibinfo{author}{Wiscombe, W.~J.}
\newblock \bibinfo{journal}{\bibinfo{title}{{Improved Mie scattering
  algorithms}}}.
\newblock {\emph{\JournalTitle{Applied optics}}} \textbf{\bibinfo{volume}{19}},
  \bibinfo{pages}{1505--1509} (\bibinfo{year}{1980}).

\bibitem{Haberle2019DocumentationCycle}
\bibinfo{author}{Haberle, R.~M.} \emph{et~al.}
\newblock \bibinfo{journal}{\bibinfo{title}{{Documentation of the NASA/Ames
  Legacy Mars Global Climate Model: Simulations of the present seasonal water
  cycle}}}.
\newblock {\emph{\JournalTitle{Icarus}}} \textbf{\bibinfo{volume}{333}},
  \bibinfo{pages}{130--164} (\bibinfo{year}{2019}).

\bibitem{Mayer2012LibRadtranGuide}
\bibinfo{author}{Mayer, B.}, \bibinfo{author}{Kylling, A.},
  \bibinfo{author}{Emde, C.}, \bibinfo{author}{Hamann, U.} \&
  \bibinfo{author}{Buras, R.}
\newblock \bibinfo{journal}{\bibinfo{title}{{libRadtran user’s guide}}}.
\newblock {\emph{\JournalTitle{Edition for libRadtran version}}}
  \textbf{\bibinfo{volume}{1}} (\bibinfo{year}{2012}).

\bibitem{Rees1989PhysicsAtmosphere}
\bibinfo{author}{Rees, M.~H.}
\newblock \emph{\bibinfo{title}{{Physics and chemistry of the upper
  atmosphere}}} (\bibinfo{publisher}{Cambridge University Press},
  \bibinfo{year}{1989}).

\bibitem{Dahlback1991ATwilight}
\bibinfo{author}{Dahlback, A.} \& \bibinfo{author}{Stamnes, K.}
\newblock \bibinfo{journal}{\bibinfo{title}{{A new spherical model for
  computing the radiation field available for photolysis and heating at
  twilight}}}.
\newblock {\emph{\JournalTitle{Planetary and Space Science}}}
  \textbf{\bibinfo{volume}{39}}, \bibinfo{pages}{671--683}
  (\bibinfo{year}{1991}).

\bibitem{Hanna2006}
\bibinfo{author}{Hanna, M.~C.} \& \bibinfo{author}{Nozik, A.~J.}
\newblock \bibinfo{journal}{\bibinfo{title}{{Solar conversion efficiency of
  photovoltaic and photoelectrolysis cells with carrier multiplication
  absorbers}}}.
\newblock {\emph{\JournalTitle{Journal of Applied Physics}}}
  \textbf{\bibinfo{volume}{100}}, \bibinfo{pages}{74510}
  (\bibinfo{year}{2006}).

\bibitem{Doscher2014SunlightDevices}
\bibinfo{author}{D{\"{o}}scher, H.}, \bibinfo{author}{Geisz, J.~F.},
  \bibinfo{author}{Deutsch, T.~G.} \& \bibinfo{author}{Turner, J.~A.}
\newblock \bibinfo{journal}{\bibinfo{title}{{Sunlight absorption in
  water–efficiency and design implications for photoelectrochemical
  devices}}}.
\newblock {\emph{\JournalTitle{Energy {\&} Environmental Science}}}
  \textbf{\bibinfo{volume}{7}}, \bibinfo{pages}{2951--2956}
  (\bibinfo{year}{2014}).

\bibitem{Hu2013AnSystems}
\bibinfo{author}{Hu, S.}, \bibinfo{author}{Xiang, C.},
  \bibinfo{author}{Haussener, S.}, \bibinfo{author}{Berger, A.~D.} \&
  \bibinfo{author}{Lewis, N.~S.}
\newblock \bibinfo{journal}{\bibinfo{title}{{An analysis of the optimal band
  gaps of light absorbers in integrated tandem photoelectrochemical
  water-splitting systems}}}.
\newblock {\emph{\JournalTitle{Energy {\&} Environmental Science}}}
  \textbf{\bibinfo{volume}{6}}, \bibinfo{pages}{2984--2993}
  (\bibinfo{year}{2013}).

\bibitem{Hannan2018State-of-the-artRecommendations}
\bibinfo{author}{Hannan, M.~A.}, \bibinfo{author}{Hoque, M.~M.},
  \bibinfo{author}{Hussain, A.}, \bibinfo{author}{Yusof, Y.} \&
  \bibinfo{author}{Ker, P.~J.}
\newblock \bibinfo{journal}{\bibinfo{title}{{State-of-the-art and energy
  management system of lithium-ion batteries in electric vehicle applications:
  Issues and recommendations}}}.
\newblock {\emph{\JournalTitle{Ieee Access}}} \textbf{\bibinfo{volume}{6}},
  \bibinfo{pages}{19362--19378} (\bibinfo{year}{2018}).

\bibitem{Eftekhari2017EnergyResearch}
\bibinfo{author}{Eftekhari, A.}
\newblock \bibinfo{journal}{\bibinfo{title}{{Energy efficiency: a critically
  important but neglected factor in battery research}}}.
\newblock {\emph{\JournalTitle{Sustainable Energy {\&} Fuels}}}
  \textbf{\bibinfo{volume}{1}}, \bibinfo{pages}{2053--2060}
  (\bibinfo{year}{2017}).

\bibitem{Anderson2018}
\bibinfo{author}{Anderson, M.~S.}, \bibinfo{author}{Ewert, M.~K.} \&
  \bibinfo{author}{Keener, J.~F.}
\newblock \bibinfo{title}{{Life support baseline values and assumptions
  document}}.
\newblock \bibinfo{type}{Tech. Rep.} (\bibinfo{year}{2018}).

\bibitem{Do2016AnAnalysis}
\bibinfo{author}{Do, S.}, \bibinfo{author}{Owens, A.}, \bibinfo{author}{Ho,
  K.}, \bibinfo{author}{Schreiner, S.} \& \bibinfo{author}{de~Weck, O.}
\newblock \bibinfo{journal}{\bibinfo{title}{{An independent assessment of the
  technical feasibility of the Mars One mission plan–Updated analysis}}}.
\newblock {\emph{\JournalTitle{Acta Astronautica}}}
  \textbf{\bibinfo{volume}{120}}, \bibinfo{pages}{192--228}
  (\bibinfo{year}{2016}).

\bibitem{Wheeler2003CropProject}
\bibinfo{author}{Wheeler, R.~M.} \emph{et~al.}
\newblock \bibinfo{title}{{Crop production for advanced life support
  systems-observations from the Kennedy Space Center Breadboard Project}}.
\newblock \bibinfo{type}{Tech. Rep.}, \bibinfo{institution}{NASA Ames Research
  Center}, \bibinfo{address}{Mountain View, CA} (\bibinfo{year}{2003}).
\newblock \doiprefix\url{NASA/TM-2003-211184}.

\bibitem{Kleinhenz2017AnVehicle}
\bibinfo{author}{Kleinhenz, J.~E.} \& \bibinfo{author}{Paz, A.}
\newblock \bibinfo{title}{{An ISRU propellant production system for a fully
  fueled Mars Ascent Vehicle}}.
\newblock In \emph{\bibinfo{booktitle}{10th Symposium on Space Resource
  Utilization}}, \bibinfo{pages}{423} (\bibinfo{year}{2017}).

\bibitem{Owens2015BenefitsMars}
\bibinfo{author}{Owens, A.}, \bibinfo{author}{Do, S.}, \bibinfo{author}{Kurtz,
  A.} \& \bibinfo{author}{Weck, O.~d.}
\newblock \bibinfo{title}{{Benefits of additive manufacturing for human
  exploration of Mars}}.
\newblock In \emph{\bibinfo{booktitle}{45th International Conference on
  Environmental Systems}} (\bibinfo{publisher}{45th International Conference on
  Environmental Systems}, \bibinfo{year}{2015}).

\bibitem{10.3389/fspas.2021.711550}
\bibinfo{author}{Berliner, A.~J.} \emph{et~al.}
\newblock \bibinfo{journal}{\bibinfo{title}{{Towards a Biomanufactory on
  Mars}}}.
\newblock {\emph{\JournalTitle{Frontiers in Astronomy and Space Sciences}}}
  \textbf{\bibinfo{volume}{8}}, \bibinfo{pages}{120},
  \doiprefix\url{10.3389/fspas.2021.711550} (\bibinfo{year}{2021}).

\bibitem{DiProfio2009ComparisonEvaluations}
\bibinfo{author}{Di~Profio, P.}, \bibinfo{author}{Arca, S.},
  \bibinfo{author}{Rossi, F.} \& \bibinfo{author}{Filipponi, M.}
\newblock \bibinfo{journal}{\bibinfo{title}{{Comparison of hydrogen hydrates
  with existing hydrogen storage technologies: Energetic and economic
  evaluations}}}.
\newblock {\emph{\JournalTitle{International Journal of Hydrogen Energy}}}
  \textbf{\bibinfo{volume}{34}}, \bibinfo{pages}{9173--9180}
  (\bibinfo{year}{2009}).

\bibitem{Barthelemy2017HydrogenPerspectives}
\bibinfo{author}{Barth{\'{e}}l{\'{e}}my, H.}, \bibinfo{author}{Weber, M.} \&
  \bibinfo{author}{Barbier, F.}
\newblock \bibinfo{journal}{\bibinfo{title}{{Hydrogen storage: recent
  improvements and industrial perspectives}}}.
\newblock {\emph{\JournalTitle{International Journal of Hydrogen Energy}}}
  \textbf{\bibinfo{volume}{42}}, \bibinfo{pages}{7254--7262}
  (\bibinfo{year}{2017}).

\end{thebibliography}

\end{document}